\documentclass{aa}  

\usepackage{graphicx}
\usepackage{subfig}
\usepackage{fixltx2e}
\usepackage{txfonts}
\begin{document}

   \title{First direct detection of a polarized companion outside of a resolved circumbinary disk around CS\,Cha\thanks{Based on observations performed with VLT/SPHERE under program ID  098.C-0760(B) and 099.C-0891(B) and VLT/NACO under program ID 298.C-5054(B) and 076.C-0292(A)}}

   %\subtitle{Direct detection of a circumplanetary disk?}

   \author{C. Ginski\inst{1,2}
	  \and M. Benisty\inst{3,4}
	  \and R.~G. van Holstein\inst{1}
	  \and A. Juh{\'a}sz\inst{5}
	  \and T.~O.~B. Schmidt\inst{6}
	  \and G. Chauvin\inst{3,4}
	  \and J. de Boer\inst{1}
	  \and M. Wilby\inst{1}
	  \and C.~F. Manara\inst{7}
	  \and P. Delorme\inst{4}
	  \and F. M\'enard\inst{4}
	  \and P. Pinilla\inst{8}
	  \and T. Birnstiel\inst{9}
	  \and M. Flock\inst{10}
	  \and C. Keller\inst{1}
	  \and M. Kenworthy\inst{1}
	  \and J. Milli\inst{4,11}
	  \and J. Olofsson\inst{12, 13}
	  \and L. P\'erez\inst{14}
	  \and F. Snik\inst{1}
	  \and N. Vogt\inst{12}
	  }

   \institute{Leiden Observatory, Leiden University, P.O. Box 9513, 2300 RA Leiden, The Netherlands\\
              \email{ginski@strw.leidenuniv.nl}
              \and Anton Pannekoek Institute for Astronomy, University of Amsterdam, Science Park 904, 1098 XH Amsterdam, The Netherlands
              \and Unidad Mixta Internacional Franco-Chilena de Astronomia, CNRS/INSU UMI 3386 and Departamento de Astronomia, Universidad de Chile, Casilla 36-D, Santiago, Chile
              \and Univ. Grenoble Alpes, CNRS, IPAG, F-38000 Grenoble, France
              \and Institute of Astronomy, University of Cambridge, Madingley Road, Cambridge, CB3 0HA, UK
              \and LESIA, Observatoire de Paris, PSL Research University, CNRS, Sorbonne Universités, UPMC Univ. Paris 06, Univ. Paris Diderot, Sorbonne Paris Cité, 5 place Jules Janssen, 92195 Meudon, France
              \and Scientific Support Office, Directorate of Science, European Space Research and Technology Centre (ESA/ESTEC), Keplerlaan 1, 2201 AZ Noordwijk, The Netherlands
              \and Department of Astronomy/Steward Observatory, The University of Arizona, 933 North Cherry Avenue, Tucson, AZ 85721, USA
              \and University Observatory, Faculty of Physics, Ludwig-Maximilians-Universit{\"a}t M{\"u}nchen, Scheinerstr. 1, 81679 Munich, Germany
              \and Max-Planck-Institut f\"ur Astronomie, K\"onigstuhl 17, 69117, Heidelberg, Germany
              \and European Southern Observatory (ESO), Alonso de Córdova 3107, Vitacura, Casilla 19001, Santiago, Chile
              \and Instituto de F\'isica y Astronom\'ia, Facultad de Ciencias, Universidad de Valpara\'iso, Av. Gran Breta\~na 1111, Valpara\'iso, Chile
              \and N\'ucleo Milenio Formaci\'on Planetaria - NPF, Universidad de Valpara\'iso, Av. Gran Breta\~na 1111, Valpara\'iso, Chile
              \and Universidad de Chile, Departamento de Astronoma, Camino El Observatorio 1515, Las Condes, Santiago, Chile
             }

   \date{Received December, 2017; accepted April, 2018}

% \abstract{}{}{}{}{} 
% 5 {} token are mandatory
 
  \abstract
   {}
   {To understand planet formation it is necessary to study the birth environment of planetary systems. 
   Resolved imaging of young planet forming disks allows us to study this environment in great detail and find signs of planet-disk interaction, as well as disk evolution.
   In the present study we aim to investigate the circumstellar environment of the spectroscopic binary T Tauri star CS\,Cha. From unresolved mid- to far-infrared photometry it is predicted that CS\,Cha hosts a disk with a large cavity. 
   In addition, SED modeling suggests significant dust settling, pointing towards an evolved disk that may show signs of ongoing or completed planet formation.}
   {We observed CS\,Cha with the high contrast imager VLT/SPHERE in polarimetric differential imaging mode to resolve the circumbinary disk in near infrared scattered light. 
   These observations were followed-up by VLT/NACO L-band observations and complemented by archival VLT/NACO K-band and HST/WFPC2 I-band data.}
   {We resolve the compact circumbinary disk around CS\,Cha for the first time in scattered light. We find a smooth, low inclination disk with an outer radius of $\sim$55\,au (at 165\,pc). 
   We do not detect the inner cavity but find an upper limit for the cavity size of $\sim$15\,au. 
   Furthermore, we find a faint co-moving companion with a projected separation of 210\,au from the central binary outside of the circumbinary disk.
   The companion is detected in polarized light and shows an extreme degree of polarization (13.7$\pm$0.4\,\%  in J-band). 
   The companion's J- and H-band magnitudes are compatible with masses of a few M$_\mathrm{Jup}$. However, K-, L- and I-band data draw this conclusion into question.
   We explore with radiative transfer modeling whether an unresolved circum-companion disk can be responsible for the high polarization and complex photometry.
   We find that the set of observations is best explained by a heavily extincted low mass ($\sim 20\,\mathrm{M}_\mathrm{Jup}$) brown dwarf or high mass planet with an unresolved disk and dust envelope.}
   {}


   \maketitle
%
%-------------------------------------------------------------------

\section{Introduction}

In the past few years high contrast and high resolution observations across a large wavelength range have revealed a variety of distinct features in planet forming disks. 
Multiple ringed systems were uncovered such as HL\,Tau (\citealt{2015ApJ...808L...3A}), HD\,97048 (\citealt{2016ApJ...831..200W}, \citealt{2016A&A...595A.112G}, \citealt{2017A&A...597A..32V}) or TW\,Hya (\citealt{2016ApJ...820L..40A}, \citealt{2017ApJ...837..132V}). 
Other systems like MWC\,758 (\citealt{2013ApJ...762...48G}, \citealt{2015A&A...578L...6B}), HD\,100453 (\citealt{2015ApJ...813L...2W}, \citealt{2017A&A...597A..42B}) or Elias\,2-27 (\citealt{2016Sci...353.1519P}) show huge spiral arms or variable shadows (HD\,135344\,B, \citealt{2017ApJ...849..143S}).
It is still unclear whether these features in general or in part are linked to ongoing planet formation, or rather to other processes within the disks.
In addition to ever more detailed images of circumstellar disks, a growing number of giant planets at wide orbital separations (typically $>$100\,au) are discovered (e.g. HD\,106906\,b, \citealt{2014ApJ...780L...4B}; HD\,203030\,b, \citealt{2006ApJ...651.1166M}; CVSO\,30\,c, \citealt{2016A&A...593A..75S}).
These objects are of particular interest to understand planet formation mechanisms, since they are the youngest planets that we have discovered and we can study their atmospheres in great detail via resolved spectroscopy. 
Yet these objects are also particularly puzzling, because typical planet formation mechanisms such as core accretion should take much longer than 100\,Myr at these distances (\citealt{1996Icar..124...62P}), while the typical dissipation timescale of gas rich disks is at least an order of magnitude shorter (\citealt{2001ApJ...553L.153H}).
Clearly, detailed characterization of other, younger, systems is required to refine the current paradigm and to understand whether the observed disk structures are linked to planet formation.
In this work we concentrate on a previously unresolved disk around a nearby T Tau object.\\
CS\,Cha is a young (2$\pm$2\,Myr, \citealt{2008ApJ...675.1375L}) classical T Tauri object of spectral type K2Ve (\citealt{1977A&A....61...21A}, \citealt{2014A&A...568A..18M}), located in the Chamaeleon I association at a distance of 165$\pm$30\,pc (combined estimate from \citealt{1997A&A...327.1194W}, \citealt{1999A&A...352..574B} following \citealt{2008A&A...491..311S}).\footnote{We note that in a recent study by \cite{2017arXiv171004528V}, the distance to the Cha\,I cloud was estimated to be slightly larger at 179\,pc. This is well covered by our uncertainties and we prefer to use the smaller distance for better comparability with previous studies until a direct distance measurement for CS\,Cha by Gaia becomes available.}
\cite{2007A&A...467.1147G} found that CS Cha is likely a single lined spectroscopic binary with a minimum mass of the secondary component of 0.1\,M$_\odot$ and a minimum orbital period of 2482\,d ($\sim$4\,au semi-major axis, assuming a system mass of 1\,M$_\odot$).
In a later study by \cite{2012ApJ...745..119N} the binary nature of CS\,Cha was confirmed. They could fit the broadened spectral lines with two Gaussian profiles, making the system potentially a double lined spectroscopic binary. They found a flux ratio of the two components of 1.0$\pm$0.4.\\
CS\,Cha is well known to feature a large infrared excess in its spectral energy distribution (SED) with a pronounced dip at 10\,$\mu$m (see e.g. \citealt{1992ApJ...385..217G}). 
The lack of emission at this wavelength regime was attributed to a large cavity by several studies (\citealt{1992ApJ...385..217G}, \citealt{2007ApJ...664L.111E}, \citealt{2009ApJ...700.1017K}, \citealt{2011ApJ...728...49E}, \citealt{2016MNRAS.458.1029R}), indicating that the system might be in a transition stage from a young gas-rich disk to a debris disk.
The radius of the cavity has been a subject of intense modeling using unresolved photometric measurements. \cite{2007ApJ...664L.111E, 2011ApJ...728...49E} find rather large cavity radii between 38\,au and 43\,au, while a more recent study by \cite{2016MNRAS.458.1029R} based on Herschel data estimates a smaller radius of 18$^{+6}_{-5}$au.
The most likely explanation is that the disk cavity is caused entirely by the stellar binary companion, since the cavity size is within a factor of a few of the binary semi-major axis.
ALMA band 3 observations by \cite{2016ApJ...823..160D} did not resolve the disk with a beam size of 2.7$\times$1.9\,arcsec, limiting the outer extent of the disk to radii smaller than 169\,au for the population of mm-sized dust grains.\\
Radiative transfer modeling of the unresolved photometry by \cite{2007ApJ...664L.111E} suggested that significant dust settling and large dust grains (5\,$\mu$m) are needed to fit the SED in the far infrared and mm wavelength ranges.
This hints at an advanced stage of dust evolution.
\cite{2014ApJ...795....1P} note that they resolve circumstellar structure around CS\,Cha with 3.3\,cm ATCA observations outside of 30\,arcsec. Since it can be excluded that this emission stems from the disk itself, they conclude that it is likely a jet, which is launched from the disk at a position angle of $\sim162^\circ$.\\
We used the SPHERE (Spectro-Polarimetric High-contrast Exoplanet REsearch, \citealt{2008SPIE.7014E..18B}) extreme adaptive optics imager to study the circumstellar environment of CS\,Cha in polarized near infrared light.
Our goals were to resolve the disk cavity for the first time and to study potential features of dust evolution or planet disk interaction such as rings/gaps and spiral arms.
In addition to our SPHERE observations we used archival high contrast data to strengthen our conclusions. 

\section{Observations and data reduction}

\subsection{The initial SPHERE polarimetric observations}

\begin{figure}
\centering
\includegraphics[width=0.95\columnwidth]{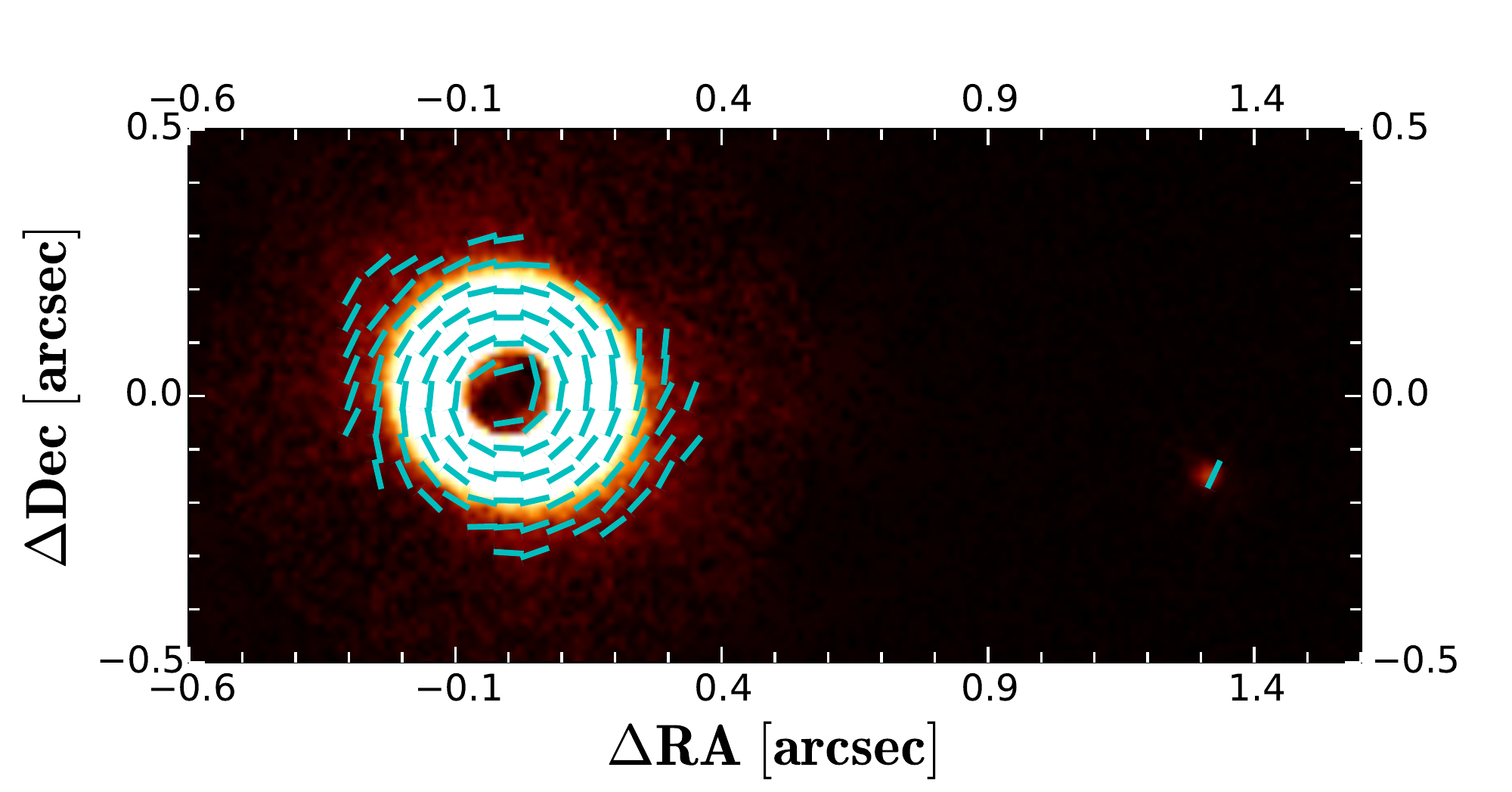}
\caption[]{Polarized intensity image of CS\,Cha and its companion in J-band, after instrumental polarization correction. 
The circumbinary disk as well as the companion are well detected in polarized light. We overlayed the angle of the linear polarized light with light blue bars. 
The companion deviates by $\sim$20$^\circ$ from an azimuthal polarization w.r.t. CS\,Cha indicating that it is intrinsically polarized and does not just scatter stellar light. }
\label{pol-vector}
\end{figure}

CS Cha was first observed with SPHERE/IRDIS (Infra-Red Dual Imaging and Spectrograph, \citealt{2008SPIE.7014E..3LD}) in Differential Polarization Imaging mode (DPI, \citealt{2014SPIE.9147E..1RL}) in J-band on February 17th 2017 as part of our ongoing program to understand dust evolution in transition disks via the distribution of small dust particles.
Conditions during the night were excellent with clear sky and an average seeing in the optical of 0.6\,arcsec and a coherence time of $\sim$5\,ms.\\ 
The (unresolved) central binary was placed behind a coronagraph with a diameter of 185\,mas (\citealt{2009A&A...495..363M}, \citealt{2011ExA....30...39C}). We used an integration time of 96\,s in individual exposures and one exposure per half wave plate (HWP) position.
A total of 11 polarimetric cycles were recorded with a combined integration time of 70.4\,min. In addition to the science data, we recorded star center frames at the beginning and end of the sequence as well as flux calibration frames and sky frames. 
For the star center frames a symmetrical waffle pattern is induced on the deformeable mirror that produces 4 satellites spots in the image. These spots can be used to accurately determine the position of the source behind the coronagraph (\citealt{2013aoel.confE..63L}).
For the flux frames the central source was offset from the coronagraph and a total of 10 images were taken with an individual exposure time of 2\,s and a neutral density filter in place in order to prevent saturation.\\
The data reduction follows generally the description given in \cite{2016A&A...595A.112G}, with the main difference being the instrumental polarization and cross-talk correction. We give a short summary here.
Our reduction approach uses the double difference method (\citealt{2001ApJ...553L.189K}, \citealt{2004IAUS..221..307A}). 
For this purpose we first subtract the ordinary and the extraordinary beam to create individual Q$^+$, Q$^-$, U$^+$ and U$^-$ images, corresponding to HWP positions of 0$^\circ$,45$^\circ$, 22.5$^\circ$ and 67.5$^\circ$.
We then subtract Q$^-$ and Q$^+$ (and U$^-$ and U$^+$) to remove the instrumental polarization downstream from the HWP within SPHERE. This is done on a cycle by cycle basis before all resulting images are median combined to obtain the Stokes Q and U images.
We also create a total intensity image, i.e. Stokes I, from our data. This is done by adding in all cases ordinary and extraordinary beams and then median combining all resulting images over all polarimetric cycles.
The Stokes Q and U images still contain residual instrumental polarization mainly induced by the VLT/UT3 mirror 3 and SPHERE mirror 4. To most accurately determine the angles and degree of linear polarization, it is necessary to correct for the instrumental polarization and cross-talk.
For this purpose we used the detailed Mueller matrix model and correction method of van Holstein et al., in prep. (including telescope mirrors, instrument common path and IRDIS itself).
This model was calibrated using an unpolarized standard star as well as the SPHERE/IRDIS internal (polarized) calibration light source and was validated with polarimetric observations of the TW\,Hya disk.
The correction was performed on each individual double difference image taking the rotation angles of all optical components from the image headers into account.
The instrument polarization model was successfully applied in several recent studies of circumstellar disks imaged with SPHERE, such as the cases of T\,Cha (\citealt{2017A&A...605A..34P}), DZ\,Cha (\citealt{2018A&A...610A..13C}) and TWA\,7 (\citealt{2018arXiv180401929O}), as well as for the observation of substellar companion polarization in the case of the HR\,8799 system (\citealt{2017arXiv170907519V}).\\
Finally we used the Stokes Q and U images to compute the radial Stokes parameters Q$_\phi$ and U$_\phi$ (see \citealt{2006A&A...452..657S}). The Q$_\phi$ image contains all azimuthally polarized flux as positive signal and radially polarized flux as negative signal.
U$_\phi$ contains all flux with polarization angles 45$^\circ$ offset from radial or azimuthal directions. In the case of single scattering by a central source, we expect all signal to be contained in Q$_\phi$ and thus U$_\phi$ can be used as a convenient noise estimate.
This is typically a valid assumption for disks seen under low inclination (\citealt{2015A&A...582L...7C}).
We show our final reduced polarimetric images in Fig.~\ref{disk-main}.  
We show the total intensity image in Fig~\ref{companion-main}.\\
In our polarimetric images we clearly detected a compact, low inclination circumstellar disk in scattered light around CS\,Cha. Furthermore, we detected in our total intensity images a faint companion candidate approximately 1.3\,arcsec to the West of CS\,Cha.
After inspection of the polarized intensity images at the companion position, it became apparent that the companion is also detected in polarized light. 
We show the final polarized intensity image including circumstellar disk and companion overlaid with the angle of linear polarization in Fig~\ref{pol-vector}.
Since the companion was detected in polarized light as well as in total intensity we can calculate its degree of linear polarization. We discuss this in detail in section \ref{comp: pol-degree}.

\subsection{Archival NACO imaging data}

The CS Cha system was previously observed with VLT/NACO (\citealt{2003SPIE.4841..944L}, \citealt{2003SPIE.4839..140R}) as part of a stellar and sub-stellar multiplicity survey among young Chamaeleon members (see \citealt{2012A&A...546A..63V} for results of that survey). Observations were carried out on February 17th 2006, i.e. exactly 11 years before our new SPHERE observations.
The data was taken in standard jitter mode in the Ks-band. Integration time for each individual exposure was 1\,s, and 35 exposures were taken and co-added per jitter position.
The total integration time of the data set was 11.7\,min. \\
We used ESO-Eclipse for the standard data reduction of the NACO data. This consisted of flat-fielding and bad pixel masking, as well as sky subtraction.
The individual reduced images were then registered with respect to the central source and median combined.\\
In addition to standard data reduction, we removed the radial symmetric part of the stellar PSF by subtracting a 180$^\circ$ rotated version of the image from itself. 
This was done in order to highlight faint companions at close angular separations and enable an accurate photometric and astrometric measurement without influence of residual stellar flux. The final reduced images are shown in Fig.~\ref{companion-main}. 
We re-detected the faint companion candidate first seen in our SPHERE observations in the final reduced NACO image.

\subsection{Archival HST/WFPC2 observations}

In addition to the NACO archival observations, CS\,Cha was also observed with the Hubble Space Telescope's Wide Field and Planetary Camera 2 (HST/WFPC2, \citealt{1994AAS...184.2402T}) on February 18th 1998.
CS\,Cha was centered in the Planetary Camera sub-aperture of WFPC2 with an effective pixel scale of 46\,mas/pixel. The observations consisted of two exposures each in the F606W and the F814W filters, i.e. the WFPC2 equivalents of R and I-band.
Exposure times for the F606W filter were 8\,s for the first exposure and 100\,s for the second exposure with gain settings of 14 and 7 respectively. In the F814W filter the exposure times were 7\,s and 80\,s with the same gain setup.
The innermost 2 pixels of the primary PSF as well as additional pixels along the central pixel readout column were saturated in all exposures. The data was reduced using the standard archival HST/WFPC2 pipeline.\\
To increase the detectability of faint point sources around the primary star we subtracted a scaled reference star PSF from the long exposure images. As reference star we used the K5 star HD\,17637, which was imaged for that purpose in the same program as the science data.
As noted by \cite{2000ApJ...538..793K}, the two main factors in achieving a good PSF subtraction result are a similar spectral type of the reference star and science target, as well as the placement of the reference star on the detector.
Due to the under-sampling of the HST PSF, it is important to use a reference star that was imaged as close in detector position as possible to the science target. From the multiple images that were taken of HD\,17637 we thus chose the one with the smallest angular separation from the position of the science target in both filters.
Since the reference star and CS\,Cha both had a saturated PSF core, we could not use the PSF peak for scaling of the reference star PSF. We instead used an annulus along the unsaturated flanks of the PSF to compute the scaling factor.\\
After subtraction of the reference star, we detected a faint point source at the expected companion candidate position in the F814W images with a signal-to-noise ratio of 5.0. The companion candidate was hidden under one of the bright diffraction spikes of the primary PSF. 
We show the subtracted and non-subtracted image in Fig.~\ref{companion-main}. In the F606W data set we could not find a significant detection at the companion candidate position. 

\subsection{NACO L-band follow-up observations}

To image CS Cha in the thermal infrared, we used again VLT/NACO. 
The observations were acquired on April 28th 2017, using
the angular differential imaging (ADI) mode
of NACO with the L' filter and the L27 camera following the
strategy described by \cite{2012A&A...548A..33C}.
The NACO detector cube mode was in addition used for frame
selection with exposure time of 0.2\,s. A classical dithering
sequence was used with the repetition of five offset positions
to properly remove the sky contribution. In the end,
the typical observing sequence represented a total of 57 cubes
of 100 frames each, i.e a total integration time of 19 min for an
observing sequence of 45 min on target. Two sequences of non saturated
PSFs were acquired using a neutral density filter
at the beginning and the end of each observing sequence
to monitor the image quality. These data also served for
the calibration of the relative photometric and astrometric
measurements. The reduction of the ADI saturated dithered datacubes was performed
with the dedicated pipeline developed at the Institut
de Plan\'{e}tologie et d’Astrophysique de Grenoble (IPAG; \citealt{2012A&A...548A..33C}) providing
various flavors of ADI algorithms. At the separation of the candidate, the background noise
is the main source of limitation. Spatial filtering
and simple derotation or classical ADI are therefore sufficient to process the ADI data.\\
After final data reduction we did not detect the companion candidate in our NACO L$_p$-band data.

\subsection{SPHERE polarimetric follow-up observations}

\begin{figure*}
\centering
\includegraphics[width=0.95\textwidth]{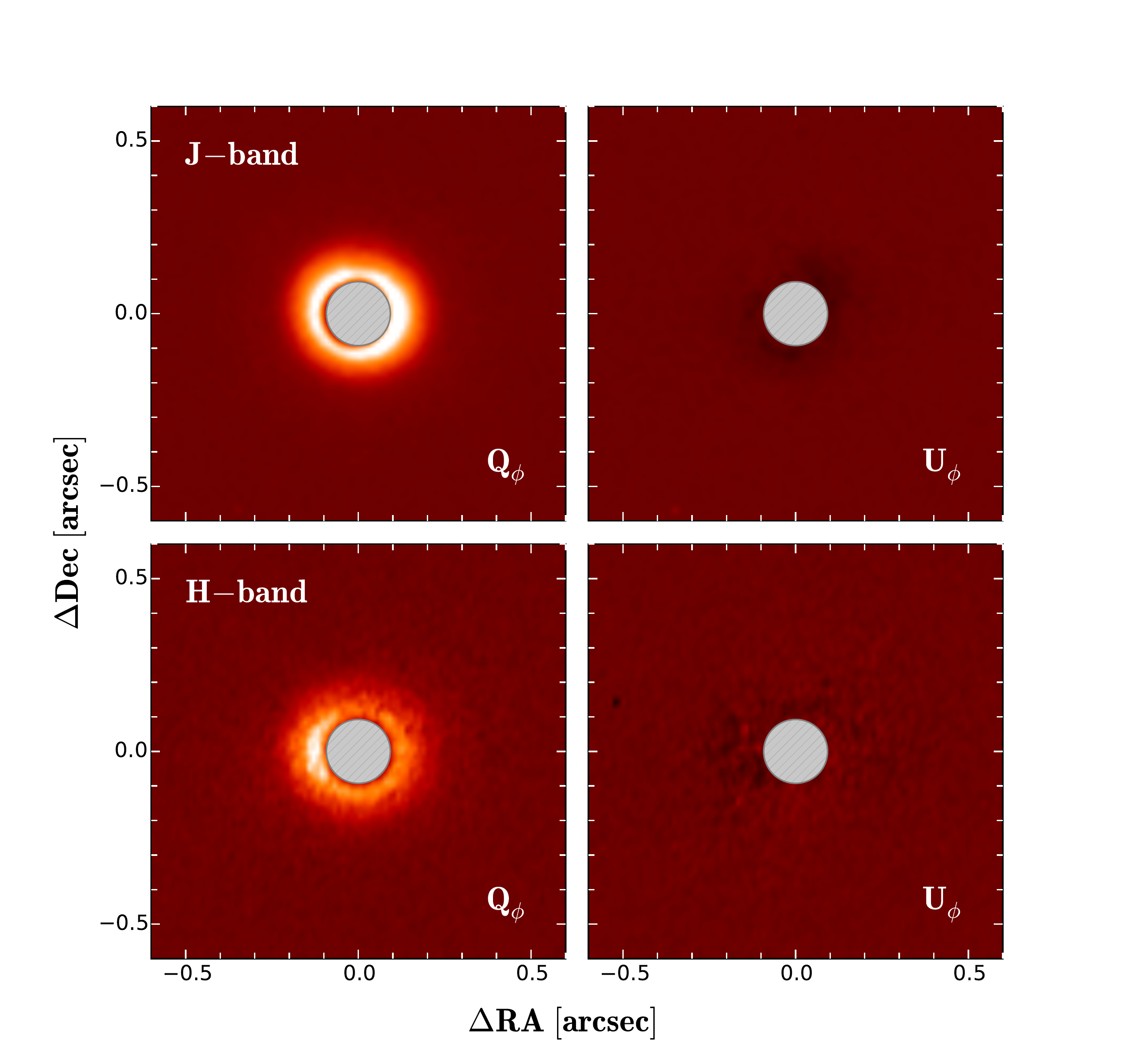}
\caption[]{\textit{1st row:} Reduced SPHERE DPI J-band Q$_\phi$ and U$_\phi$ as well as intensity image. North is up and East to the left.  
\textit{2nd Row:} The same for our H-band observations. Color scale (linear) and stretch are the same for all Q$_\phi$ and U$_\phi$ images. We did not correct for the 1/r$^2$ drop-off in stellar irradiation. 
The grey hatched disk overplotted on the images shows the size of the utilize coronagraph.
}
\label{disk-main}
\end{figure*}

\begin{figure*}
\centering
\includegraphics[width=0.98\textwidth]{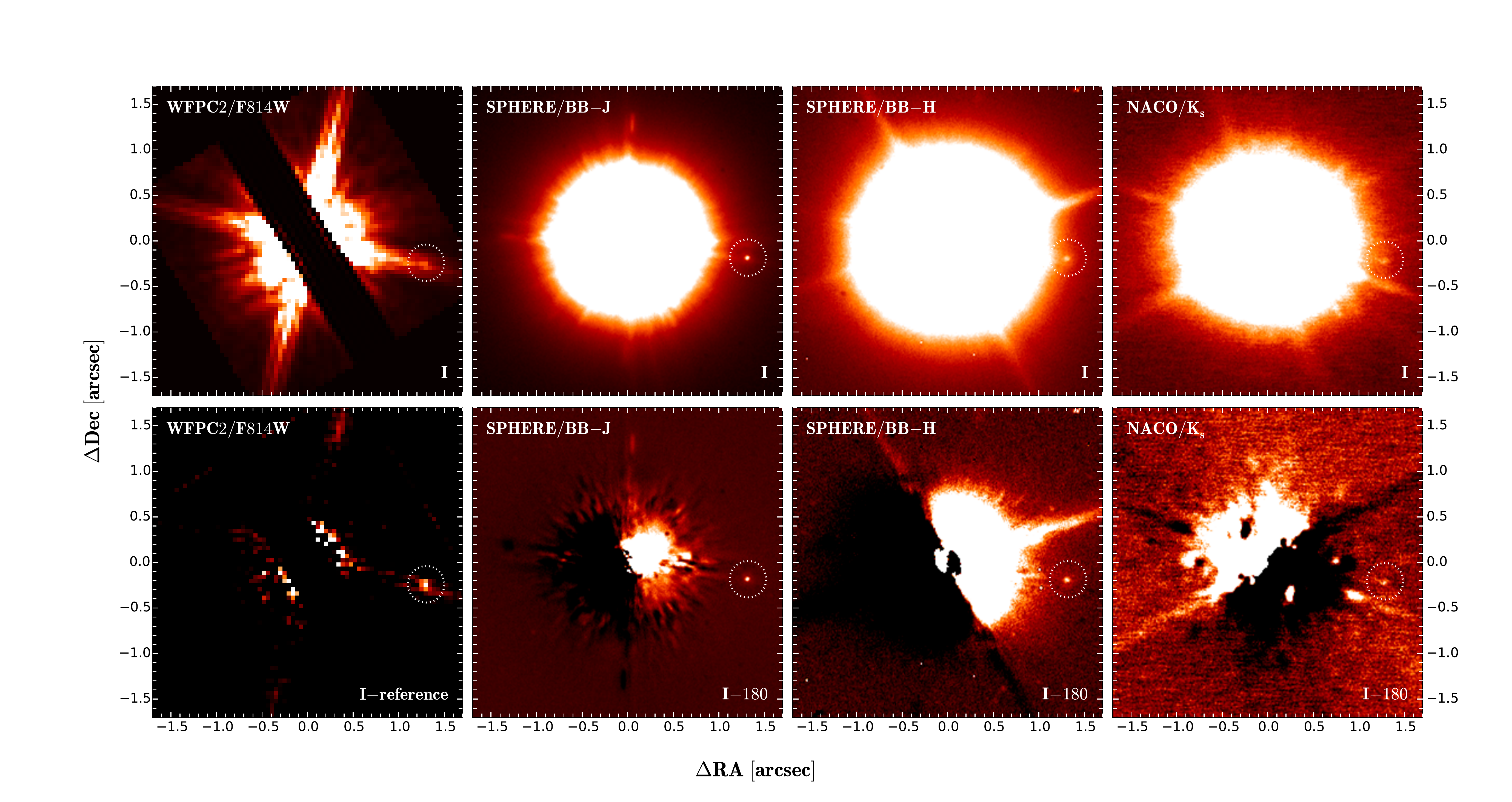}
\caption[]{\textit{1st row:} SPHERE J and H-band intensity images as well as the NACO K$_{\mathrm{s}}$-band image and the WFPC2 F814W image of CS\,Cha. North is up and East to the left. In all images the position of the faint companion candidate is marked by a white dashed circle.  
\textit{2nd Row:} The same images as above but subtracted with a 180$^\circ$ rotated version of themselves to remove the bright stellar halo. In the case of WFPC2 we subtracted a reference star scaled to the flux of CS\,Cha to remove the bright stellar PSF and especially the bright diffraction spike on top of the companion position.
In the WFPC2 images we removed the central columns containing the PSF peak since they were heavily saturated.}
\label{companion-main}
\end{figure*}

Our initial SPHERE observations were followed-up on June 18th 2017 with SPHERE/IRDIS DPI observations in H-band with the goal to obtain H-band photometry of the companion candidate and to confirm its detection in polarized light. The conditions during the observations were overall poor. Even though the seeing in the optical was on average only 0.7\,arcsec, the coherence time was very short, on the order of 2\,ms on average during the observations. 
This lead to a much poorer AO correction compared to the previous J-band observations.\\
The setup of the June observations was similar to the first set of observations in February. The central binary was again placed behind the 185\,mas coronagraph. We used a slightly shorter individual exposure time of 64\,s due to the unstable weather conditions.
We recorded a total of 7 polarimetric cycles with a combined integration time of 29.9 minutes. \\
Data reduction was performed analogously to the previous J-band data set. 
We found that the circumstellar disk and the companion candidate were again detected in polarized light. We could also detect the companion in our stacked total intensity images. 
Final reduced images are shown in Fig.~\ref{disk-main} and Fig.~\ref{companion-main}.

\section{Astrometric confirmation of the companion}

\begin{figure*}
\centering
\subfloat[][PM from NOMAD catalogue]{
\includegraphics[scale=0.35]{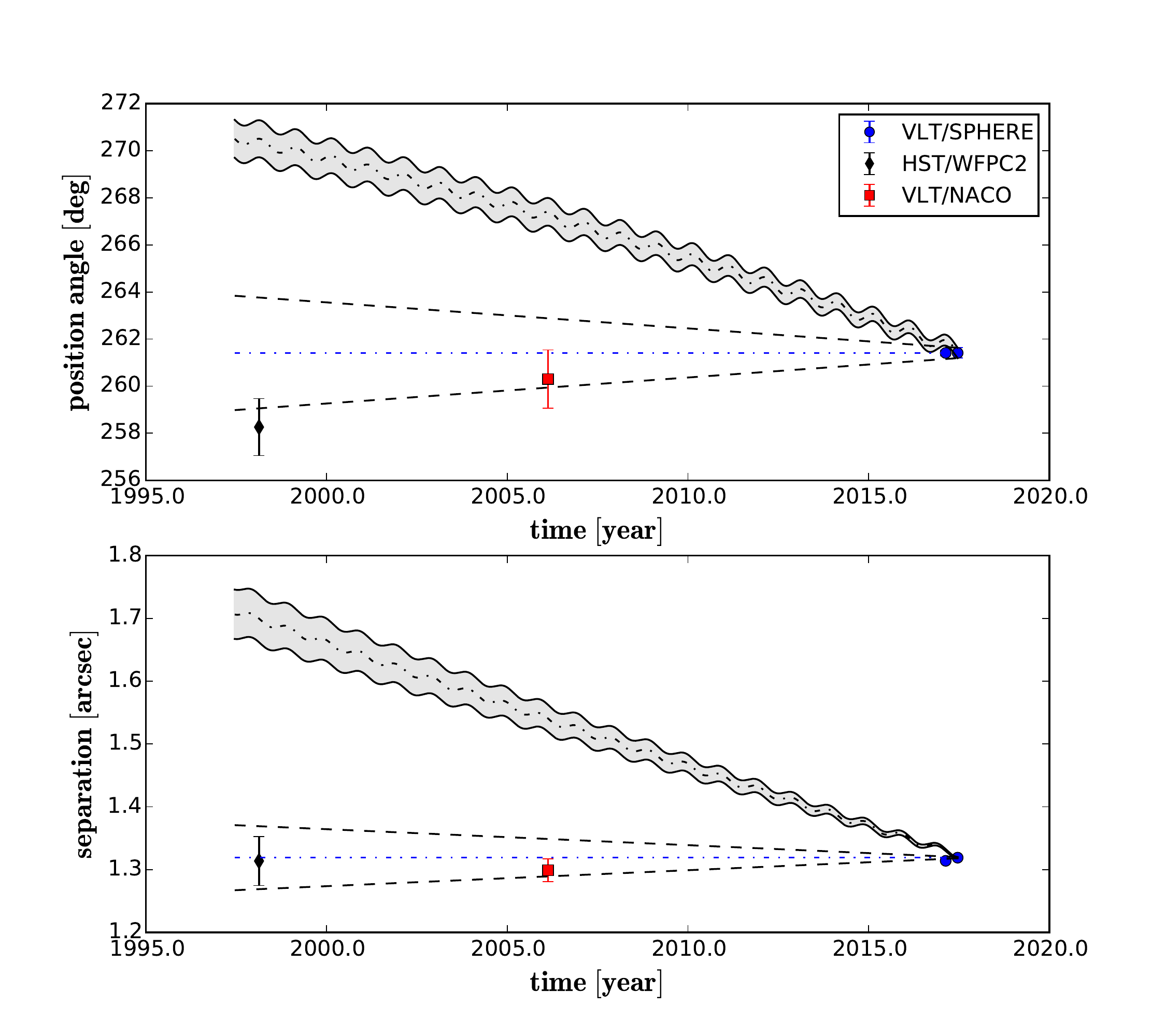}
\label{pm-nomad}
}
\subfloat[][PM from SPM4 catalogue]{
\includegraphics[scale=0.35]{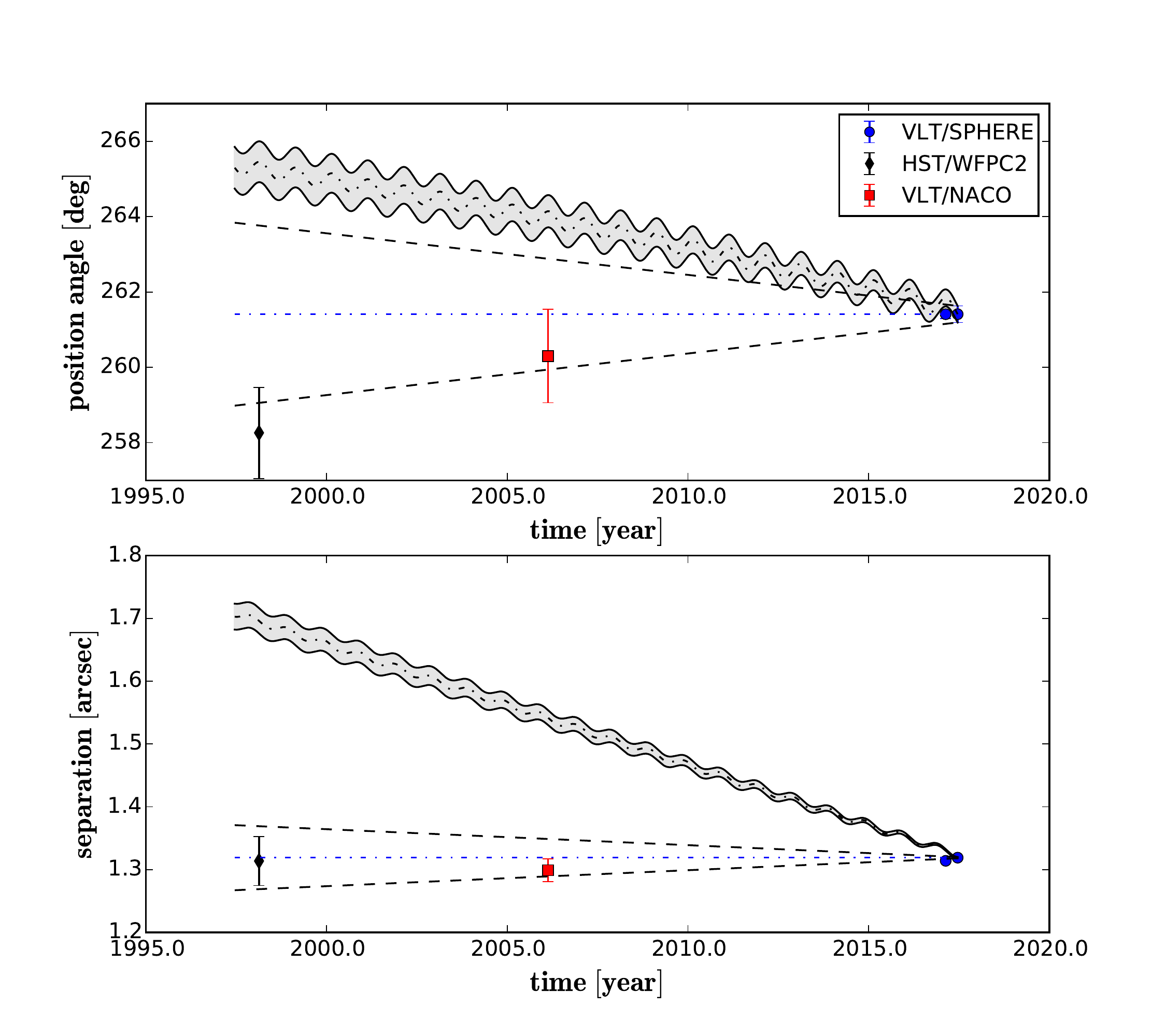}
\label{pm-spm4}
}

\subfloat[][PM from Smart et al. 2014]{
\includegraphics[scale=0.35]{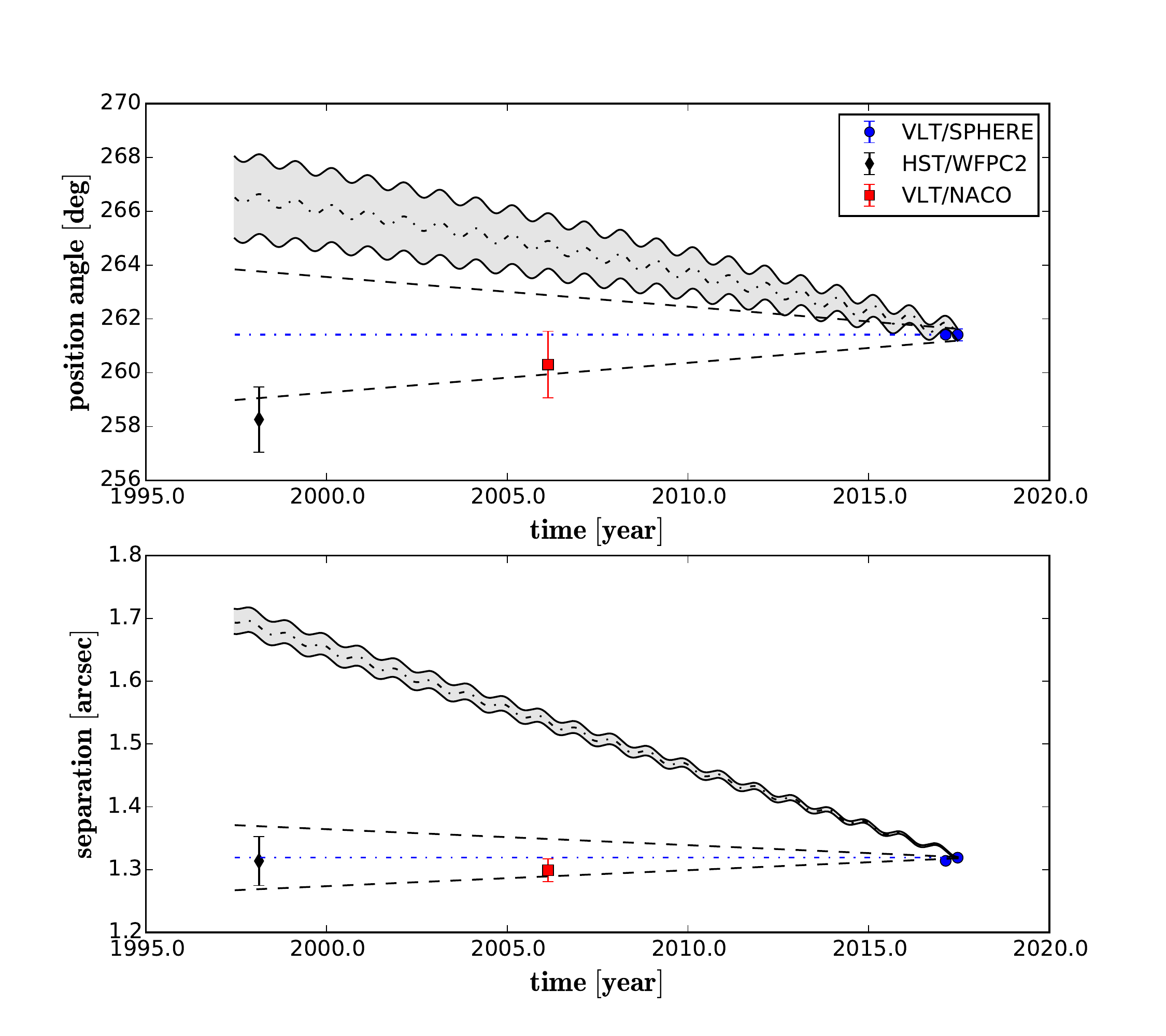}
\label{pm-smart}
}
\caption[]{Proper motion diagrams of the companion relative to CS\,Cha. The wobbled gray lines are the area in which a non-moving background object would be expected. The "wobbles" are due to the parallactic displacement of such an object visible during the Earths revolution around the Sun.
The dashed lines mark the area in which a co-moving companion would be located. The dashed lines take potential orbital motion into account assuming an inclination (circular face-on for the position angle and circular edge-on for the separation) and total system mass (1\,M$_\odot$, i.e. this assumes that the mass of the companion is small compared to CS\,Cha).
In all three diagrams the companion is co-moving with CS\,Cha and thus in all likelihood gravitationally bound. We note that we see a small differential motion in position angle across our 19\,year observation baseline, which is consistent with a circular face-on (or close to face-on) orbit.} 
\label{pm-plots} 
\end{figure*}

Since the companion was detected in our new SPHERE data as well as the archival NACO and HST data, we were able to to test if the companion is co-moving with CS\,Cha.
To ensure minimal contamination by the central stars' flux in the SPHERE and NACO images, we subtracted a 180$^\circ$ rotated version of the images from the original (see Fig.~\ref{companion-main}). 
For the HST F814W image we used the reference star subtracted image to determine the companion position. \\
In the case of the SPHERE and NACO images we used IDL \texttt{starfinder} (\citealt{2000A&AS..147..335D}) to fit a reference PSF to the companion and extract its position in detector coordinates.
As reference PSF we used the unsaturated stellar primary in the NACO image and the dedicated flux frames taken for the SPHERE images.
Since the separation of the companion is smaller than the average isoplanatic angle at Paranal (see e.g. \citealt{2000A&AS..144...39M}), no significant distortions of the companion PSF compared to the primary star PSF are expected.
For the HST PSF under-sampling and residuals from the diffraction spike made PSF fitting problematic. Instead we used ESO-MIDAS (\citealt{1983Msngr..31...26B}) to fit a two dimensional Gaussian to the companion position.
Measurements were repeated several times with different input parameters in terms of measuring box size and starting position to ensure that the fit converged well.
We used the average value of all measurements as the final extracted companion position. We used the individual fitting uncertainties of the Gaussian fit as the uncertainty of the companion's detector position.
We ensured that this uncertainty was significantly larger than the standard deviation of multiple repeated measurements with different initial parameters.\\
To extract the stellar position, we used different approaches for the SPHERE, NACO and HST data. For the NACO data no coronagraph was used, so we used the same approach to extract the stellar position as was used for the companion position.
However, for SPHERE no direct measurement was possible since the central source is obscured by the coronagraph. Instead we used the center calibration frames to determine stellar position, as described in \cite{2013aoel.confE..63L}.
Since we had multiple center frames taken at the beginning and end of the sequence, we used the deviation between the recovered positions as the uncertainty of the central source position measurement.
For the HST image the primary star was heavily saturated with significant column bleeding, making a fit to the remaining stellar PSF difficult.
Instead we fit linear functions to the positions of the diffraction spikes and used their intersection as stellar center position.\\
To translate the recovered detector position for the central binary and the companion into on-sky separation and position angle, our observations required an astrometric calibration.
For the archival NACO data several binary stars were imaged as astrometric calibrators during the same night as the science data as part of the original program. 
The results of these astrometric calibrations (including also potential orbital motion of the binary calibrators), are given by \cite{2012A&A...546A..63V}.\\
For the SPHERE data, calibrators are regularly imaged during the ongoing SPHERE GTO survey.
Primary calibrators are stellar clusters such as 47\,Tuc, $\Theta$\,Ori\,B and NGC\,6380. The results of these astrometric calibrations are given in \cite{2016SPIE.9908E..34M}.
In addition, detailed solutions for the geometric distortions were calculated by these authors. The instrument has proven to be extremely stable within their given uncertainties.
We thus utilize their results for the broad band J and H filter to calibrate our data. We also use their distortion solution to correct geometric distortions in our SPHERE image.
For the true north of the J-band data we use the more recent measurement published in \cite{2017A&A...605L...9C}, done within a few days of our observations. 
The uncertainties for the SPHERE data include those of the detector coordinates of central source and companion as well as the calibration uncertainty and the uncertainty of the distortion solution.
Lastly, for the HST data we used the astrometric calibration provided in the image header.
We list final results in table ~\ref{tab: astrometry}.\\
After we extracted the astrometry in all epochs, we measured the proper motion of the companion relative to CS\,Cha. The final results are shown in Fig.~\ref{pm-plots}.
We show three different diagrams, since the proper motion of CS\,Cha is given with slightly different values in the NOMAD (\citealt{2004AAS...205.4815Z}) and SPM4 (\citealt{2011AJ....142...15G}) catalogs, as well as by \cite{2014A&A...570A..87S}.
In all three cases we can clearly reject the background hypothesis with 7.1 to 8.7\,$\sigma$ in separation and with 4.4 to 8.5\,$\sigma$ in position angle. 
Within the given uncertainties we observe no significant relative motion in separation over our $\sim$19\,yr baseline. 
However we observe relative motion in position angle, which is consistent with a circular face-on or low inclination orbit, i.e. with a similar inclination as is observed for the resolved circumbinary disk.
Within the given error bars the companion is thus co-moving with the primary stars. This is a very strong indication that the companion is gravitationally bound to CS\,Cha.
In particular it is extremely unlikely that the companion is a blended extragalactic source, since such a source would have to move at very high velocity and would need to be by-chance aligned in proper motion and close to CS\,Cha.
The probability for a blended galactic source might be slightly higher. To quantify this we used the \texttt{TRILEGAL} (\citealt{2005A&A...436..895G, 2012ASSP...26..165G}) population synthesis model to compute the number of expected galactic sources in close vicinity of CS\,Cha.
As input we gave the galactic coordinates of CS\,Cha as well as the J-band magnitude of the companion as limiting magnitude. Following \cite{2012A&A...546A..10L} the number of expected objects can then be translated into a probability to find a background object at a certain separation.
Using this approach we find that the chance of a faint blended galactic source within 1.3\,arcsec of CS\,Cha is 0.4\,\%, i.e. improbable at the 2.9\,$\sigma$ level. 
Such a source would then still need to be by-chance aligned in proper motion with CS\,Cha making this scenario even less likely.
One last concern might be that the companion could be a blended local source within the Cha\,I cloud but several pc behind CS\,Cha. 
For example \cite{1998A&A...338..977C} found a number of very faint and highly embedded YSOs in Cha\,I. 
To test the likelihood of a by chance aligned local source in Cha\,I we checked the dispersion of proper motions of known members. 
As input we used the catalog by \cite{2000A&A...361.1143T}, which contains 29 such members, including CS\,Cha. We find that the dispersion in proper motion is quite high with $\sim$18\,mas/yr in right ascension and $\sim$73\,mas/yr in declination.
In contrast we find that the companion shows no significant deviation from the proper motion of CS\,Cha in right ascension and only 3.6$\pm$1.8\,mas/yr in declination, which can be well explained by orbital motion as mentioned earlier.
In Fig.~\ref{app: cambresy}, in the appendix, we furthermore show that the recovered colors of the companion do not match the YSO colors in Cha\,I by \cite{1998A&A...338..977C}.
We can thus firmly exclude a blended local object as well.\\
We overall conclude that the companion is in all likelihood gravitationally bound to CS\,Cha. We explore the orbital motion of the companion in detail in section \ref{comp:nature-discussion}.

\begin{table*}[t]
\small
 \centering
  \caption{Astrometric meassurements and calibrations of all observation epochs.}
  \begin{tabular}{@{}llcccc@{}}
  \hline   
	\hline
 	 Epoch			& Instrument 	  	& Pixel Scale [mas/pixel]	& True North [deg] 	& Separation [arcsec] 	& Position Angle [deg]\\
 	\hline
	1998.1339		& WFPC2			& 45.52$\pm$0.01		& 31.69$\pm$0.005	& 1.314$\pm$0.039	& 258.26$\pm$1.21	\\
	2006.1311		& NACO			& 13.24$\pm$0.18		& 0.18$\pm$1.24		& 1.299$\pm$0.018	& 260.30$\pm$1.24	\\
	2017.1311		& SPHERE/IRDIS		& 12.263$\pm$0.009		& -1.71$\pm$0.06	& 1.314$\pm$0.002	& 261.41$\pm$0.12	\\
	2017.4617		& SPHERE/IRDIS		& 12.251$\pm$0.009		& -1.75$\pm$0.11	& 1.319$\pm$0.001	& 261.40$\pm$0.23	\\
 	
 \hline\end{tabular}

\label{tab: astrometry}
\end{table*}

\section{Photometric measurements and detection limits}
\subsection{SPHERE and NACO photometry}
\newpage

\begin{figure}
\centering
\includegraphics[width=0.95\columnwidth]{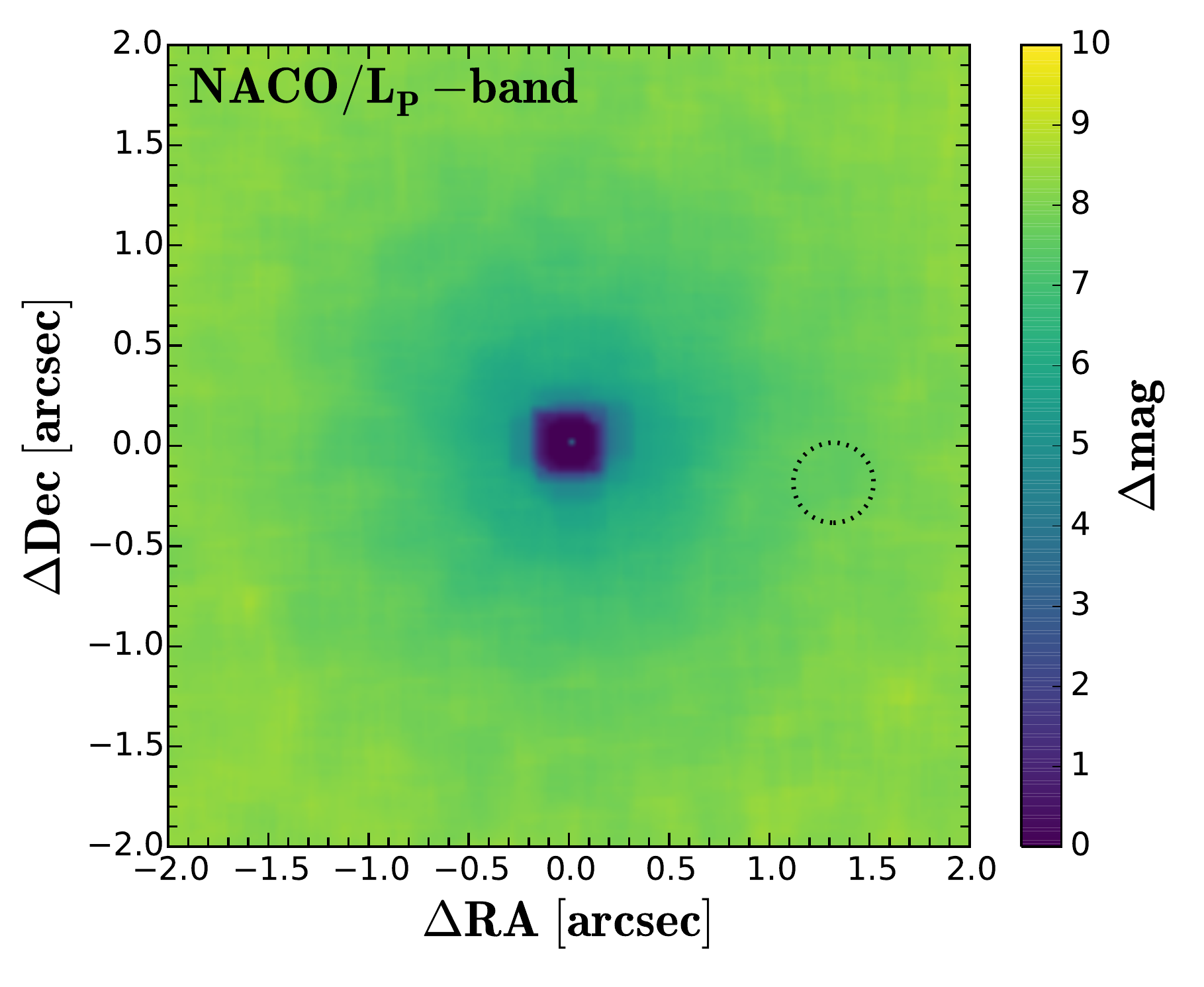}
\caption[]{Detection limit map derived from our NACO L$_p$-band observations. Detection limits are given in relative contrast to the the primary star. We mark the expected position of the companion with a black, dashed circle. The companion was not detected in these observations and should thus exhibit a contrast larger than 8.2\,mag relative to CS\,Cha.} 
\label{naco-l-limit}
\end{figure}

To understand the nature of the faint companion, we performed photometric measurements in all bands in which the companion was detected and derived upper limits for the non-detections.
In our SPHERE J and H-band epochs, as well as in the NACO K$_s$-band epoch, the companion was well detected but was still located close enough to CS\,Cha so that the background at the companion position is dominated by the bright stellar halo.
We did not image PSF reference stars (neither was a PSF reference available for the archival NACO data), thus we assumed that the low frequency structure of the stellar halo is approximately radial symmetric.
To remove this radial symmetric halo, we subtracted 180$^\circ$ rotated versions of the images from themselves. The results are shown in Fig.~\ref{companion-main}. 
While strong signal remains within $\sim$0.5\,arcsec of CS\,Cha, the companion position appears free of strong residuals.\\
After this initial background subtraction we utilized IDL \texttt{starfinder} to perform PSF fitting photometry to measure relative brightness between the companion and CS\,Cha (the latter in the unsubtracted images). We used the flux calibration frames for the SPHERE observations to obtain an unsaturated reference PSF. 
For the NACO K$_s$-band image we used CS\,Cha itself as reference PSF since it was not saturated during the science sequence. Once a PSF fitting result was obtained we subtracted the companion from the data to check for strong residuals at the companion position.
The results are given in table~\ref{tab: photometry} as differential magnitudes. To convert the differential J, H and K$_s$ magnitudes to apparent magnitudes of the companion we used the corresponding 2MASS (\citealt{2003yCat.2246....0C}) magnitudes of CS\,Cha as calibration.
We then also list absolute magnitudes for which we assumed a distance of 165$\pm$30\,pc. For conversion to physical fluxes we used a HST/STIS Vega spectrum as well as the filter curves of SPHERE and NACO. \\
We note that we find a clear systematic uncertainty in the SPHERE H-band observations, induced by the poor observing conditions.
In particular the coherence time of the atmosphere degraded during the sequence, with longer values at the start than at the end of the sequence. 
However, our flux calibration frames were only taken at the end of the sequence, i.e. in the worst observing conditions, and thus have lower Strehl than the previous science images in the sequence. 
Thus using them for the flux measurement of the companion during the whole sequence over-predicts the companion flux.   
To estimate this systematic effect we sub-divided the science sequence into four equally long bins, which we reduced individually in order to detect the companion in each bin.
We then measured the relative loss of signal in the companion due to changing weather conditions between all bins. We found a deviation of 0.46\,mag between the first and the last bin.
We consider this as an additional error term for the lower limit (since we know the direction of the effect) of the companion flux in H-band.\\
As mentioned earlier the companion was not detected in our NACO L-band observation. We thus evaluated the detection limit of our observation.
The detection performances reached by our observation were estimated
by computing 2D detection limit maps, at 5$\sigma$ in terms of L$_p$ contrast with
respect to the primary. We computed the pixel-to-pixel noise within a
sliding box of 1.5 $\times$ 1.5 FWHM. The detection limits
were then derived by taking the ADI flux loss using fake planet injection and the transmission of the
neutral-density filter into account, and were normalized by the
unsaturated PSF flux. Our final detection limits map is shown in Fig.~\ref{naco-l-limit} and the computed detection limit at the companion position is given in table~\ref{tab: photometry}.
We use the WISE (\citealt{2012wise.rept....1C}) W1 magnitude as close proxy for the L-band magnitude of the primary star to convert contrast limits to apparent and absolute magnitude limits.

\subsection{WFPC2 photometry and detection limits}

\begin{figure}
\centering
\includegraphics[width=0.95\columnwidth]{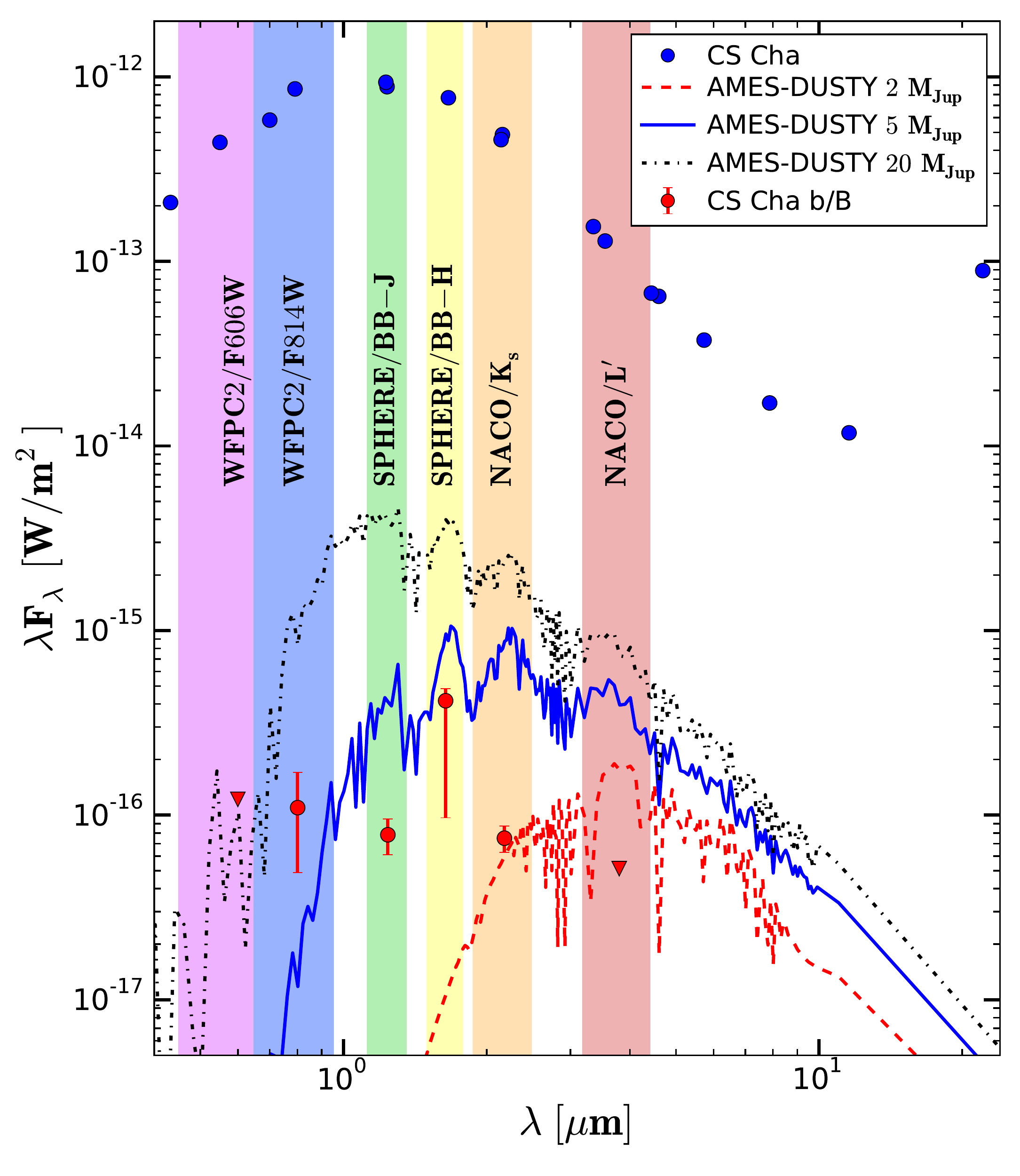}
\caption[]{Spectral energy distribution of CS\,Cha (blue dots) and its companion (red dots and triangles). Pointing down triangles denote upper limits. Spectral flux densities were computed from broad band photometry using a Vega spectrum and the broad band filter curves. All values for the companion are given in table~\ref{tab: photometry}}
\label{sed}
\end{figure}

To estimate the brightness of the companion in the WFPC2 F814W filter, several analysis steps were necessary. The primary star was saturated in the long exposure in which we detected the companion.
To enable a relative measurement of the companion brightness we thus first determined the brightness of the primary star in the exposure. 
For this purpose we used TinyTim (\citealt{2011SPIE.8127E..0JK}), a program to generate HST point spread functions based on the instrument setup, target spectral type, time of observations and position on the detector.
We created a matching PSF for the WFPC2 F814W observations and then fitted this theoretical PSF to the unsaturated flanks of the CS\,Cha PSF by application of a scaling factor.
We then used this scaled theoretical PSF for the relative brightness measurement with the companion.\\
The photometry of the companion in the WFPC2 image is challenging since it is contaminated by the bright diffraction spike of the primary star. 
Even after subtraction of a reference star, residuals of this diffraction spike are still visible around the companion position. 
Due to the low S/N of the detection and the under-sampling of the HST PSF, we decided against PSF fitting photometry in this case and instead applied aperture photometry. 
For this purpose we measured the flux of the companion in a 3$\times$3 pixel box centered on the brightest pixel of the companion PSF. 
We then estimated the local background by measurements with the same box 3 pixels moved in radial direction towards and away from the the central star along the diffraction spike.
The average of both measurements was then subtracted from the companion measurement. 
To estimate the uncertainty of the background measurement we computed the standard deviation in the background apertures and multiplied it by the surface area of the aperture.
In addition to the uncertainty of the background we took into account the read noise of the WFPC2 planetary camera for a gain setting of 7e$^-$/DN. We used a read noise of 5 e$^-$/pix.
Overall the measurement is strongly dominated by the uncertainty of the background, which is a factor 4 higher than the estimated read noise.
We give our result for the relative brightness measurement in magnitudes in table~\ref{tab: photometry}.\\
To convert this relative measurement to an apparent magnitude of the companion, we determined the Vega magnitude of CS\,Cha in the F814W filter. 
For this purpose we calculated the total flux of CS\,Cha in the F814W filter using the filter curve and the spectral energy distribution of CS\,Cha, shown in Fig.~\ref{sed}.
We then converted this to a Vega magnitude by comparison with the flux of Vega in the same filter. We give apparent and absolute magnitudes for the companion also in table~\ref{tab: photometry},
along with the physical flux of the companion in the F814W filter.\\
In the F606W filter the companion was not detected. We thus estimated detection limits at the companion position. For this purpose we measured the standard deviation at the companion position in a 3$\times$3 pixel aperture.
We again used TinyTim to create an unsaturated reference PSF scaled to the primary star brightness on the detector. 
Given the noise at the companion position and using the primary star as reference we then computed the limiting magnitudes for a 5\,$\sigma$ detection.
The result is given in table~\ref{tab: photometry}.

\section{Polarization of the companion}
\label{comp: pol-degree}

\begin{figure}
\centering
\includegraphics[width=0.9\columnwidth]{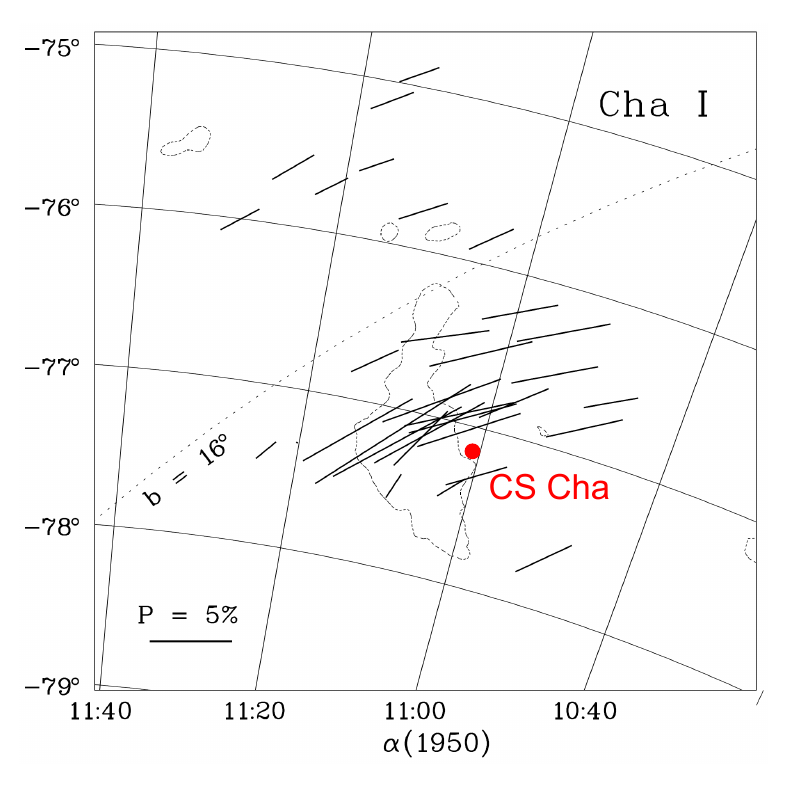}
\caption[]{Reproduction of Fig.~2 from \cite{1997A&AS..122...95C}. Shown is a visual polarization map of the Cha\,I dark cloud using 33 stars they observed. 
Average angle and degree of polarization is indicated by the solid line vector field. 
We show the position of CS\,Cha in this map to indicate the expected degree of linear polarization introduced by the Cha\,I dark cloud in the optical.}
\label{polmap}
\end{figure}

Since we detected the companion both in the total intensity and in the polarized intensity images in our SPHERE/IRDIS J and H-band epochs, we could calculate the degree of linear polarization of the companion.
For this purpose we used aperture photometry in both images in each band. We used aperture photometry over PSF fitting photometry due to the slight change in the companion PSF during the double difference steps of the DPI reduction.
We checked in J-band that PSF fitting photometry and aperture photometry give consistent results in the total intensity image. We found that the results were consistent within 0.01\,mag.
We used an aperture radius of 3\,pix in J-band, which corresponds to the full width at half maximum of 2.77\,pix as measured by fitting a Moffat profile to the stellar PSF.
In H-band we used a value of 4\,pix due to the poorer observing conditions.
As in the PSF fitting photometry, we first subtracted the radial symmetric bright stellar halo from the intensity images by rotating them 180$^\circ$ and subtracting them from themselves.
We then estimated the local background with two sub-apertures in each band. In the J-band case, the companion is slightly contaminated by a stellar diffraction spike. We thus used two sub-apertures in radial direction along this spike.
In H-band we used two azimuthal sub-apertures at the same separation from the central star and offset by a few degrees from the companion position.\\
The measurement in polarized intensity was performed in the same way as in the intensity image. 
However, the measurements were actually performed in the Stokes Q and U images rather than in the combined polarized intensity image, since all signal becomes positive in this image and thus even background noise might give a spurious polarization signal.\\
We find a degree of linear polarization in J-band of 13.7$\pm$0.4\,\% and an angle of linear polarization of 153.0$^\circ \pm$0.8$^\circ$.
Our H-band results are consistent with these measurements. We find a degree of linear polarization of 14.1$\pm$1.4\,\% and an angle of linear polarization of 154.0$^\circ \pm$2.9$^\circ$.
The uncertainties in the H-band measurements are higher due to the poorer signal-to-noise compared to the J-band data. 
In both cases the error bars are strongly dominated by measurement uncertainties (due to photon, speckle and background noise), while the instrument model allows for a factor of $\sim$10 higher accuracy.\\
We now need to investigate if the polarization of the companion is intrinsic to the object, i.e. either due to scattered light from the primary stars or a central object within the companion itself, or if it is caused by interstellar dust between Earth and the CS\,Cha system.
This is of particular importance since CS\,Cha is indeed located within close proximity or behind the Cha\,I dark cloud.
Detailed optical polarization measurements of the region have been performed by \cite{1997A&AS..122...95C}. In Fig.~\ref{polmap} we show their optical polarization map of the Cha\,I cloud region and superimpose the position of CS\,Cha.
Using the nine stars in their study located closest to the position of CS\,Cha, we find an average maximum polarization degree of 6.7$\pm$1.7\,\% at a peak wavelength of 0.65\,$\mu m$.
The average angle of polarization in the I-band that they measure for the same stars is 132.08$^\circ \pm$12.9$^\circ$.
Using Serkowski's empirical law (\citealt{1975ApJ...196..261S}) we can extrapolate the expected polarization degree in the J and H-band.

\begin{equation}
p(\lambda) = p(\lambda_\mathrm{max}) \exp \left[ -K \ln^2 (\lambda / \lambda_\mathrm{max})\right]  
\end{equation}

Wherein $p(\lambda)$ is the polarization degree at the wavelength $\lambda$. The factor K was empirically determined to be dependent on the peak wavelength of the polarization degree by \cite{1982AJ.....87..695W}.

\begin{equation}
K =  1.86 \, \lambda_\mathrm{max} - 0.1
\end{equation}

Using these relations with the average values from \cite{1997A&AS..122...95C} we find that the degree of linear polarization could lie between 5.5\,\% and 3.3\,\% for the J-band and between 3.4\,\% and 2.0\,\% for the H-band.
These degrees of polarization are significantly lower than what we find for the companion giving a first indication that the polarization of the companion is indeed intrinsic and not caused by interstellar dust.\\
To test this more rigorously we measured the degree of linear polarization of the unresolved primary stars. For this purpose we used an annulus at the bright speckle halo that marks the adaptive optics correction radius and that contains only stellar light.
We find that the primary stars have a degree of linear polarization of 0.57$\pm$0.28\,\% in J-band with an angle of polarization of 133.1$^\circ \pm$8.2$^\circ$. 
For the H-band we find similar values of 0.34$\pm$0.02\,\% and 141.1$^\circ \pm$5.4$^\circ$ for the degree and angle of linear polarization respectively.
Both of these values are consistent with a previous measurement in the optical of CS\,Cha by \cite{2000A&AS..144..285Y} who find a degree of linear polarization of 0.7\,\% (but did not provide uncertainties).
The degree of polarization that we find for the primary stars is much lower than suggested by the optical data by \cite{1997A&AS..122...95C} in combination with Serkowski's law.
It could be that the average value we assumed is not a good proxy for the cloud density at the position of CS\,Cha or that CS\,Cha is located slightly in front of the cloud.
In any case, the low degree of stellar polarization strongly suggests that the high degree of polarization found for the companion is intrinsic to the object and not caused by interstellar dust if we assume both objects are located at the same distance as suggest by their common proper motion.\\
This conclusion is additionally supported by the disagreement between the angle of polarization of the companion and that of stellar sources.
In both bands the angle of linear polarization of the stellar binary is within 1\,$\sigma$ consistent with the average polarization angle in the region as determined by \cite{1997A&AS..122...95C}.
In our data we find that the companion polarization angle deviates by $\sim22^\circ$ (2.6$\sigma$) from the stellar polarization angle in J-band and by $\sim15^\circ$ (2.4$\sigma$) in H-band.
Thus it seems again plausible that the cause for the polarization of the stellar binary and the companion are different and that the companion polarization is not caused by interstellar dust.\\
Given the angle of polarization of the companion, we can finally try to understand which is the dominating source of illumination, assuming polarization by single scattering of light.
If the companion is primarily illuminated by the central stellar binary we would expect its angle of linear polarization to be azimuthal with respect to the binary position.
The expected angle of linear polarization for azimuthally scattered light at the companion position is 171.7$^\circ \pm$0.1$^\circ$. Comparing this to the more accurate angle of linear polarization in J-band,
we find a significant deviation of 18.7$^\circ \pm$0.8$^\circ$. We can thus conclude that the origin of the polarized light is not (entirely) single-scattered light emitted by the primary stars.
It is of course still possible that the linear polarization that we measure is a superposition of scattered stellar and companion emission. 
However, given the angle of polarization we can already conclude that the companion object contains a central source massive enough that we can detect its' emission.\\
Polarization can give us important information about the structure of the atmosphere of low-mass objects, as well as their direct environment.
Polarization has indeed been measured for field brown dwarfs previously (see e.g. \citealt{2002A&A...396L..35M}, \citealt{2005ApJ...621..445Z} and \citealt{2013A&A...556A.125M}), but was not detected so far for companions to nearby stars (see e.g. \citealt{2016ApJ...820..111J}, \citealt{2017arXiv170907519V}).
This is to the best of our knowledge the first time a faint and thus likely low-mass companion to a nearby star was detected in polarized light and its degree of polarization measured.
We discuss the implications for the object in detail in section~\ref{nature-section}.

\begin{table*}[t]
\small 
 \centering
  \caption{Photometric measurements of the companion. The apparent magnitudes in J, H and K$_s$-band were calculated using the closest 2MASS magnitude of CS\,Cha as calibration. 
  The apparent magnitude in the HST F814W filter was computed using the theoretical Vega magnitude of CS\,Cha in this band given its SED. 
  The absolute magnitudes were computed from the apparent magnitudes assuming a distance of 165$\pm$30\,pc. We give the central wavelength and spectral width of all filters along with the measurements.
  Spectral flux densities were computed using the filter curves of the instruments as well as a Vega spectrum taken with HST/STIS.}
  \begin{tabular}{@{}llcccccc@{}}
  \hline   
	\hline
 	 Instrument 	&	Filter			& $\lambda_c$ [$\mu$m]	& $\Delta \lambda$ [$\mu$m]	& $\Delta$mag 		& app. magnitude 	& abs. magnitude	& F$_\lambda$ [W$\cdot$m$^{-2}\cdot\mu$m$^{-1}$]\\
 	\hline
 	HST/WFPC2	&	F606W			& 0.5997		& 0.1502			& $>$8.9		& $>$20.4		& $>$14.3		& $<2.03\cdot10^{-16}$ \\
	HST/WFPC2	&	F814W			& 0.8012		& 0.1539			& 9.81$\pm$0.48		& 19.71$\pm$0.48	& 13.62$\pm$0.62	& $(1.37\pm0.76)\cdot10^{-16}$ \\ 
	SPHERE		&	BB-J			& 1.245			& 0.240				& 10.05$\pm$0.21	& 19.16$\pm$0.21	& 13.07$\pm$0.45	& $(6.31\pm1.39)\cdot10^{-17}$ \\
	SPHERE		&	BB-H			& 1.625			& 0.290				& 9.20$^{+0.61}_{-0.15}$& 17.65$^{+0.62}_{-0.16}$& 11.56$^{+0.74}_{-0.43}$	& $(2.54^{+0.41}_{-1.95})\cdot10^{-16}$ \\
	NACO		&	Ks			& 2.18			& 0.35				& 9.21$\pm$0.16		& 17.40$\pm$0.16	& 11.32$\pm$0.43	& $(3.44\pm0.56)\cdot10^{-17}$ \\
	NACO		&	Lp			& 3.80			& 0.62				& $>$8.2		& $>$16.4		& $>$10.3		& $<1.35\cdot10^{-17}$\\
 	
 \hline\end{tabular}

\label{tab: photometry}
\end{table*}

\section{The circumbinary disk around CS\,Cha}

\subsection{Position angle and inclination}

As visible in Fig.~\ref{disk-main}, we resolve for the first time a small disk around the central stellar binary in the CS\,Cha system. The disk appears compact, smooth and close to face-on.
From our scattered light images we can extract the orientation of the disk. For this purpose we measure the disk diameter in radial disk profiles with orientations between 0$^\circ$ and 360$^\circ$ in steps of 2$^\circ$.
The resulting disk diameter versus disk orientation data was fitted with the corresponding value for an ellipse. The disk diameter was defined in our radial profiles as the separation between the two outermost points at which the disk flux reaches a certain threshold.
To determine this threshold we measured the standard deviation of the background outside of the disk signal and set the threshold to a multiple of this standard deviation. In practice we found that there is a small dependency on the threshold value and the recovered disk orientation.
We thus used multiples between 5 and 100 in steps of 2 and considered the recovered median values for disk inclination and position angle, and the standard deviation between these values, as the uncertainty of our measurement.
Assuming a radial symmetric disk that only appears elliptical due to its relative inclination towards us, we find a inclination of 24.2$^\circ \pm$3.1$^\circ$ and a position angle of 75.6$^\circ \pm$2.2$^\circ$ from our J-band observation.
This disk position angle is well consistent with the position angle of the suspected jet emission of $\sim162^\circ$ detected by \cite{2014ApJ...795....1P}, since the jet position angle should be offset by 90$^\circ$ from the disk major axis.
The H-band observation has much lower signal-to-noise than the J-band observation and suffers from convolution with a rather distorted PSF (see Fig.~\ref{app: sphere_stellar}). 
We find an inclination of 34.9$^\circ \pm$10.6$^\circ$ and a position angle of 86.1$^\circ \pm$2.2$^\circ$ for this data set. The $\sim$10$^\circ$ larger position angle can be explained by the elongated PSF shape and orientation of this observation.
We thus consider the J-band measurements as final values for inclination and position angle.

\subsection{Inner and outer radius}
\label{disk-radius}

\begin{figure*}
\centering
\subfloat[][]{
\includegraphics[scale=0.4]{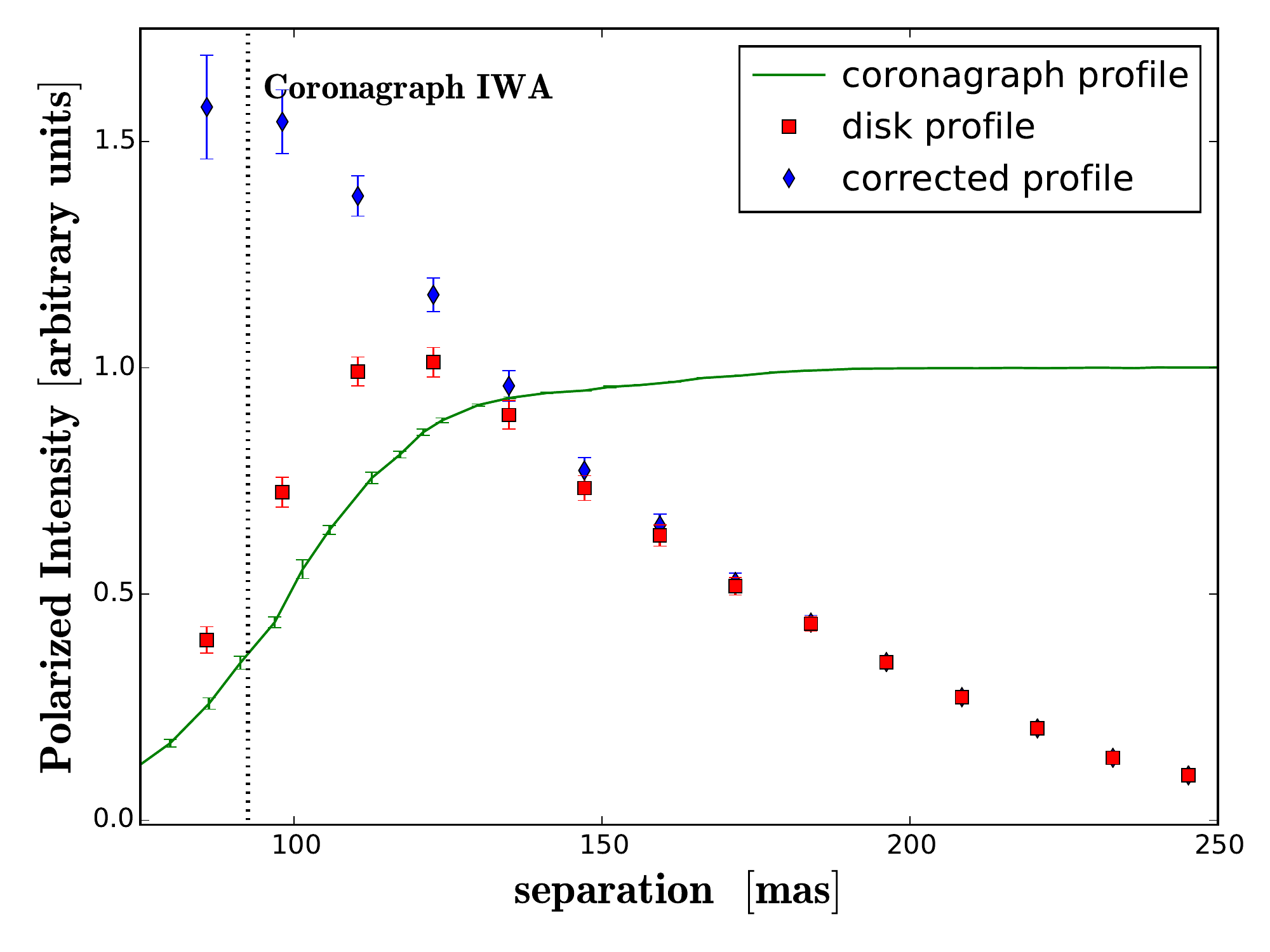}
\label{profile1}
}
\subfloat[][]{
\includegraphics[scale=0.4]{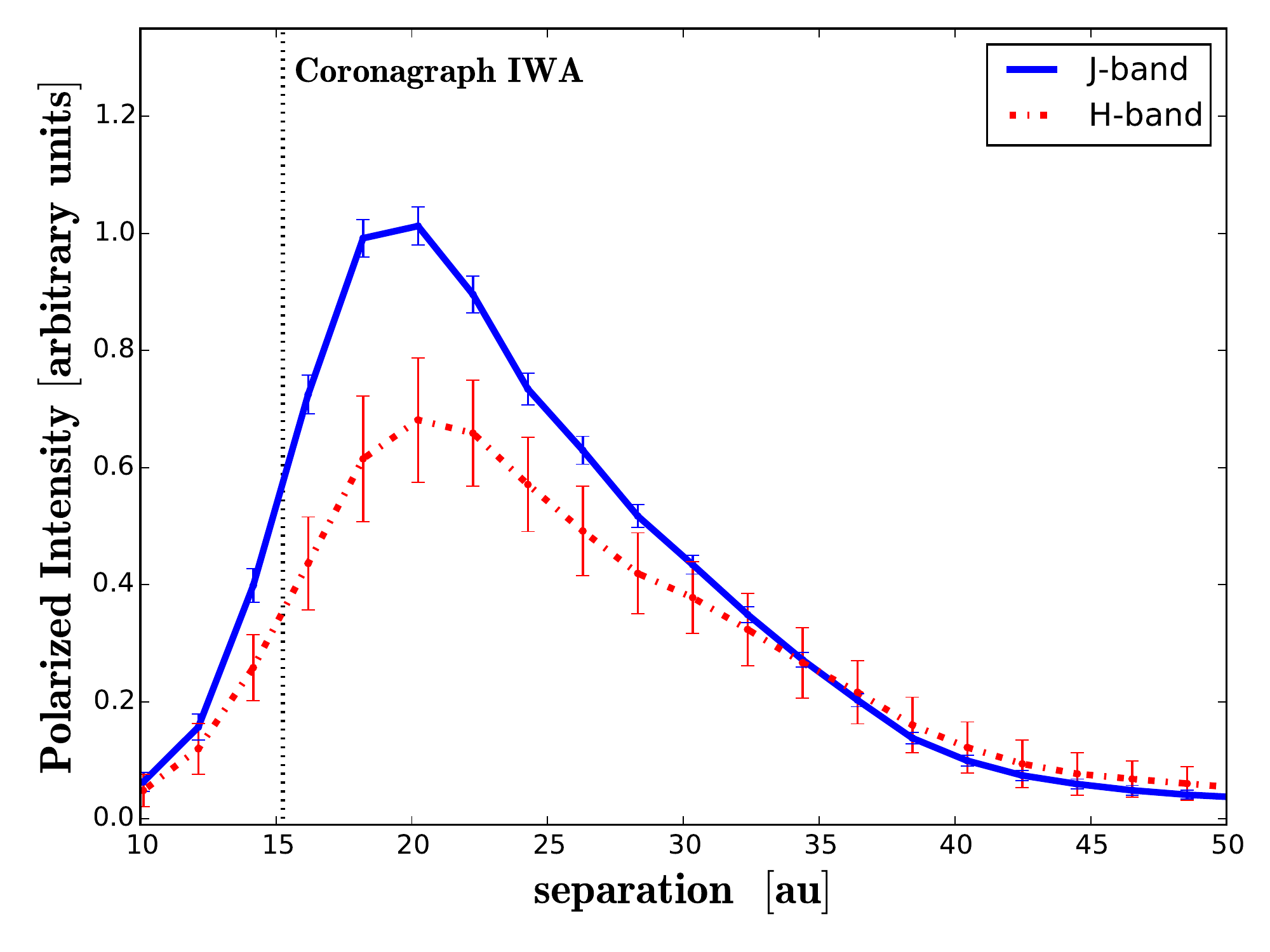}
\label{profile2}
}
\caption[]{\textit{Left:} Azimuthal average of the polarized intensity profile of the circumbinary disk around CS\,Cha in J-band (red squares). The profile was measured in the Q$_\phi$ image, while the estimated uncertainties were determined in the U$_\phi$ image.
We indicate the radius of the coronagraphic mask with the black dotted line. In addition, we show the throughput curve of the utilized coronagraph as discussed in section \ref{disk-radius} (green solid line). 
Finally we show the azimuthal disk profile corrected by the coronagraph throughput (blue diamonds).
\textit{Right:} Azimuthal average of the polarized intensity profile of the circumbinary disk around CS\,Cha in J-band (blue solid line) and H-band (red dash-dotted line). Angular separations were converted to projected separations using the distance of 165\,pc.} 
\label{disk: profile}
\end{figure*}

To measure the outer radius of the disk we considered a radial profile along the major axis as determined in the previous section. 
We then computed the radial extent at which the disk signal is for the first time 5$\sigma$ above the image background value. We again used the J-band images, due their higher quality.
We found an outer radius of scattered light of 337\,mas, i.e. 55.6\,au at a distance of 165\,pc. This is consistent with the upper limit of 169\,au given by \cite{2016ApJ...823..160D} from their unresolved ALMA observations.
Note that we are only tracing small dust at the disk surface, so that it is possible that the disk has a larger size but is partially self shadowed.
Another possibility is that the disk outer extent is larger, but that it is below the noise floor in our images due to the 1/r$^2$ drop-off of the stellar irradiation.\\ 
We show an azimuthally averaged radial profile of the disk in Fig.~\ref{disk: profile}. In this profile a decline in brightness inside of $\sim$115\,mas is visible. To investigate if this is a tentative detection of a cavity, we compared the radial disk profile with a model profile of the coronagraph attenuation.
The NIR APLC coronagraph normalization profile was calculated based on IRDIS DB\_H23 dual-band imaging observations of the 0.6" diameter disk of Ceres, performed on the 14th of December 2016. 
This was carried out in the N\_ALC\_YJH\_S coronagraph imaging mode and the Ceres disk was nodded off-center by 490 mas to provide a non-coronagraphic reference. This was used to produce a 2D attenuation profile of the coronagraph for an extended, incoherent source. 
Monochromatic Fourier modeling of the three-plane APLC coronagraph was also performed, using the APO1 SPHERE amplitude apodiser, ACL2 (185 mas diameter) focal-plane mask and NIR Lyot stop including dead actuator masks (\citealt{2011ExA....30...59G}, \citealt{2016JATIS...2b5003S}). 
This model confirmed that the observed Ceres attenuation profile is nearly diffraction-limited and azimuthally symmetric. The radial profile outside of 85 mas is dominated by direct throughput of the target, while that inside 85 mas is dominated by internally scattered light in the instrument (for full results see Wilby et al., in prep.). 
The close agreement between the forward model and observed data allows the H23-band profile to be extrapolated to J-band via an equivalent model at 1.26\,$\mu$m. This was then used to correct the radial CS Cha profile for coronagraph attenuation.\\
As visible in Fig.~\ref{disk: profile}, after the correction with the coronagraph throughput profile, no significant decline in flux is visible outside the coronagraphic mask. 
We can thus put an upper limit on the size of the inner cavity of the CS Cha disk of 15.3\,au (92.5\,mas at 165\,pc) from the scattered light imaging (tracing small dust grains). 

\section{The nature of the companion}
\label{nature-section}

To understand the nature of this new companion, we compare its SED to known substellar objects in Chamaeleon as well as theoretical model atmospheres.  
We then use the astrometry over a 19\,yr base line to determine if it is possible to constrain the companion mass from the orbital motion.
Finally we use our own radiative transfer models to explain the photometry and degree of linear polarization of the companion.

\subsection{A planetary mass object on a wide orbit?}
\label{comp:nature-discussion}

\begin{figure}
\centering
\includegraphics[width=0.95\columnwidth]{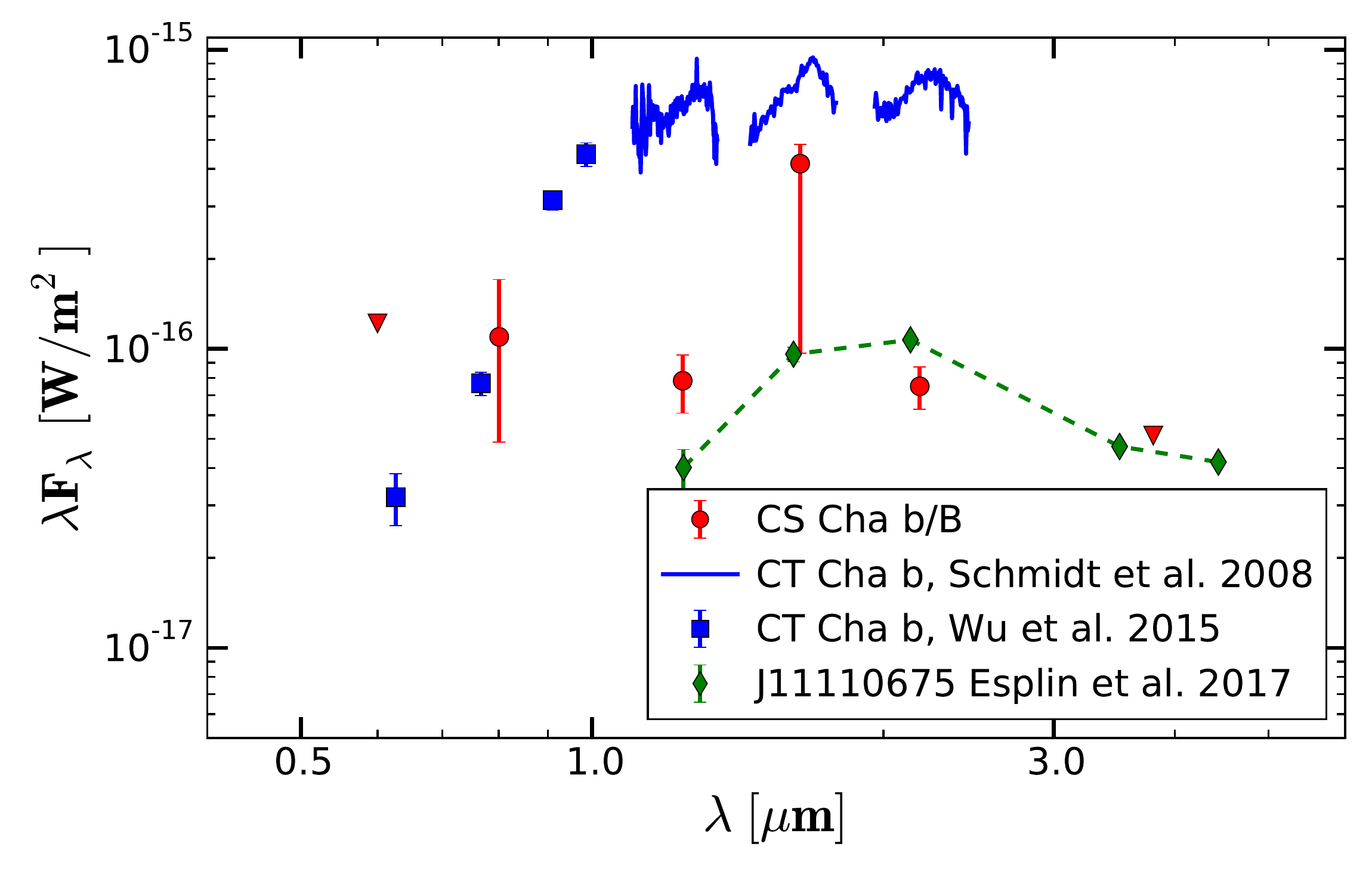}
\caption[]{Spectral energy distribution of the CS\,Cha companion (red dots and triangles). Pointing down triangles denote upper limits. We show the known substellar companion CT\,Cha\,b (\citealt{2008A&A...491..311S}, \citealt{2015ApJ...801....4W}) as well as the free floating planetary mass object in Chamaeleon Cha\,J11110675-7636030 (\citealt{2017AJ....154...46E}) as comparison.}
\label{sed-small}
\end{figure}

\begin{figure*}
\centering
\subfloat[][]{
\includegraphics[scale=0.3]{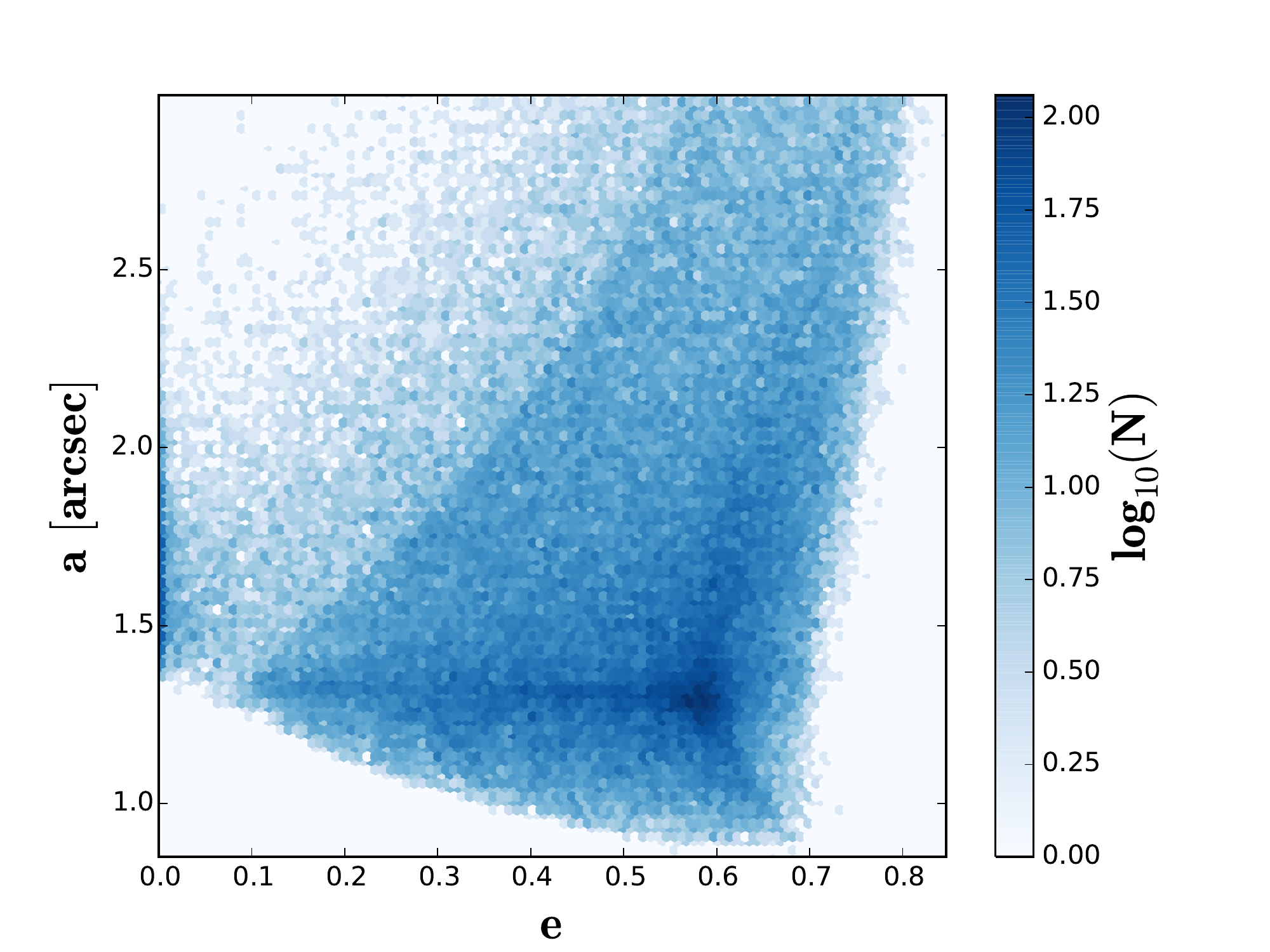}
\label{a_vs_e}
}
\subfloat[][]{
\includegraphics[scale=0.3]{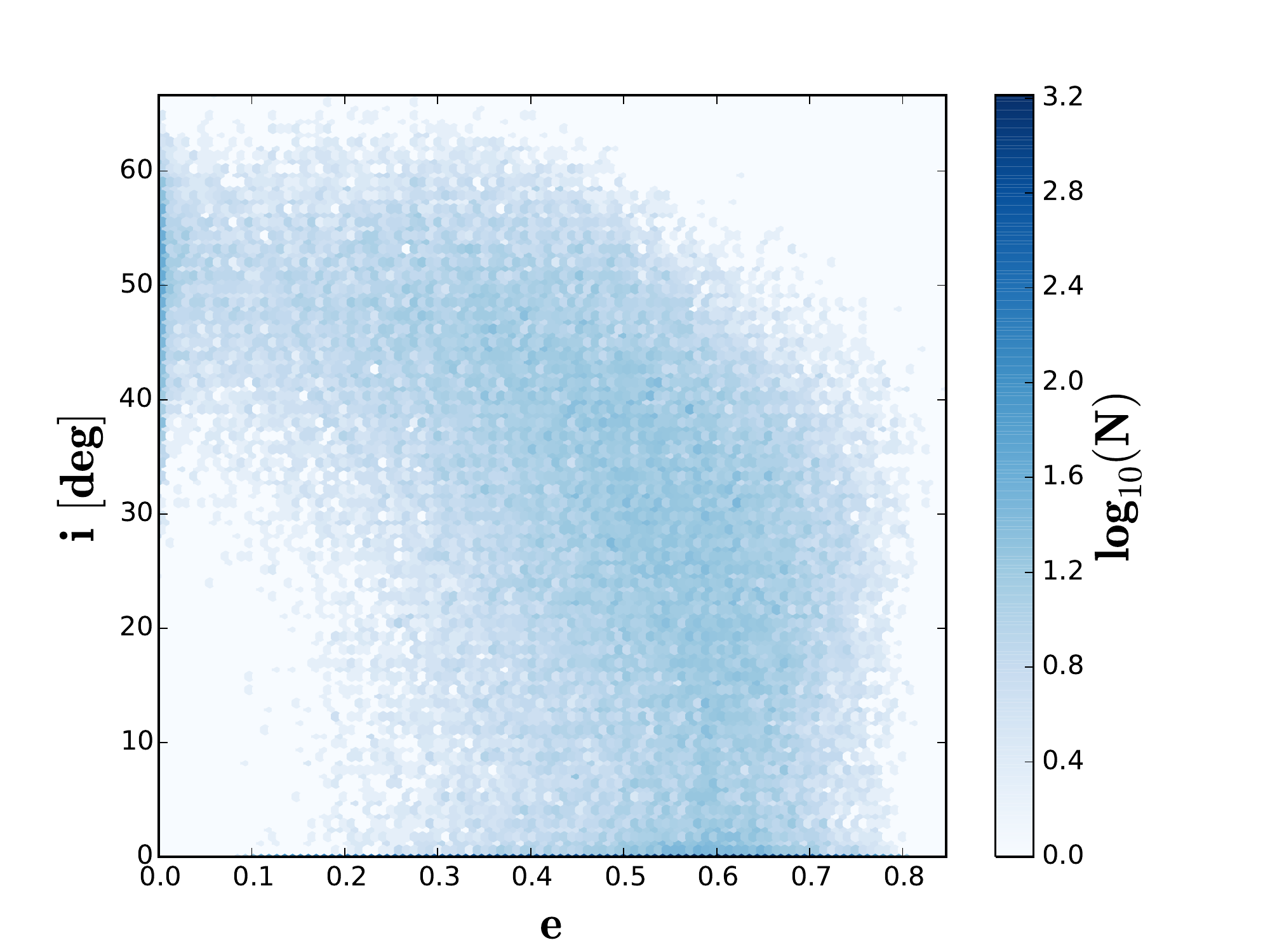}
\label{i_vs_e}
}
\subfloat[][]{
\includegraphics[scale=0.3]{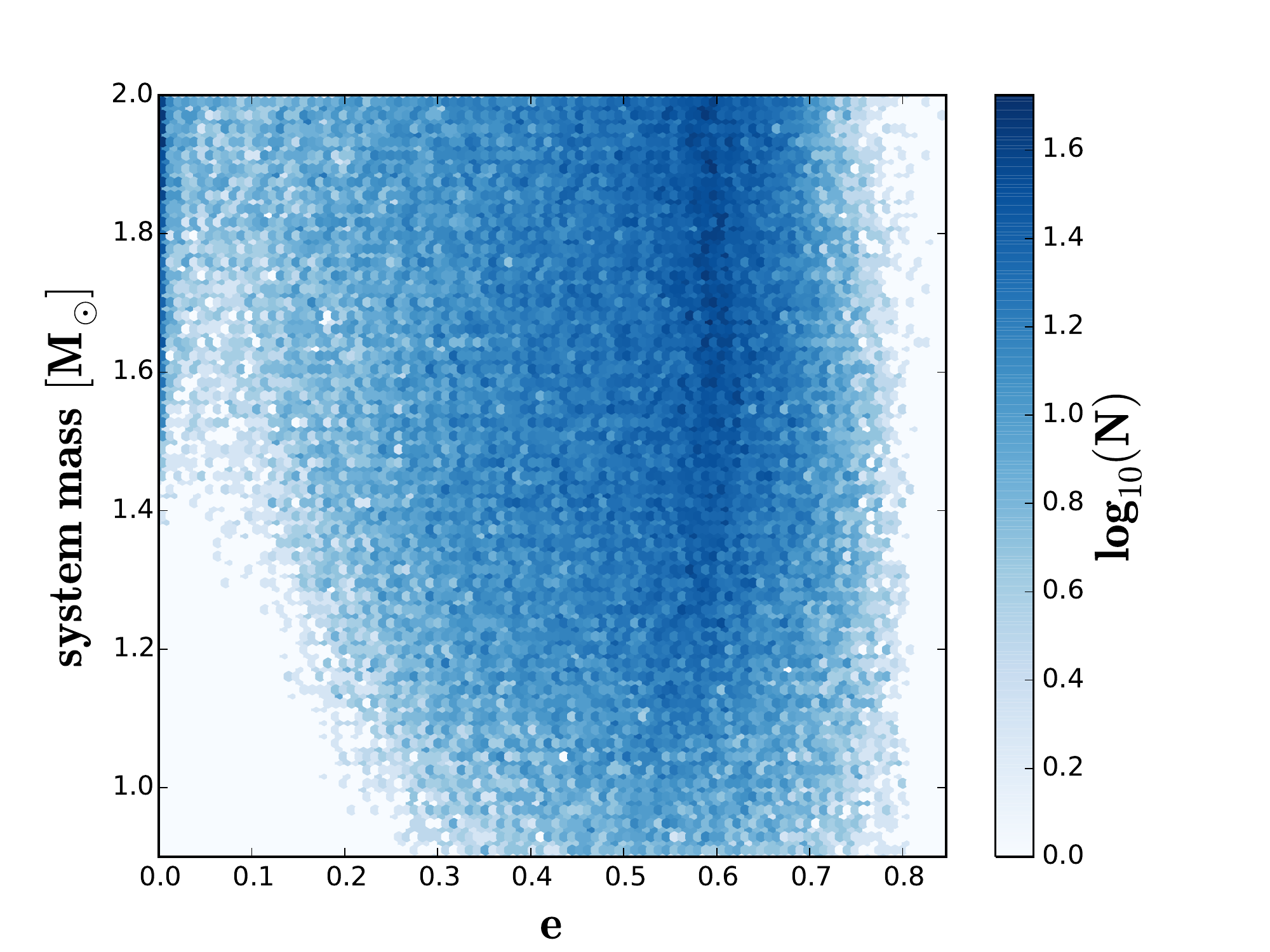}
\label{e_vs_mass}
}
\caption[]{\textit{Left:} Semi-major axis versus eccentricity distribution of all recovered orbit solutions for the companion following the LSMC approach. Shown are the 1\% best fitting orbits.
\textit{Middle:} Same as left, but for eccentricity versus inclination of the orbital plane.
\textit{Right:} Same as left, but for eccentricity versus total system mass.} 
\label{companion:lsmc}
\end{figure*}

\begin{figure}
\centering
\includegraphics[width=0.95\columnwidth]{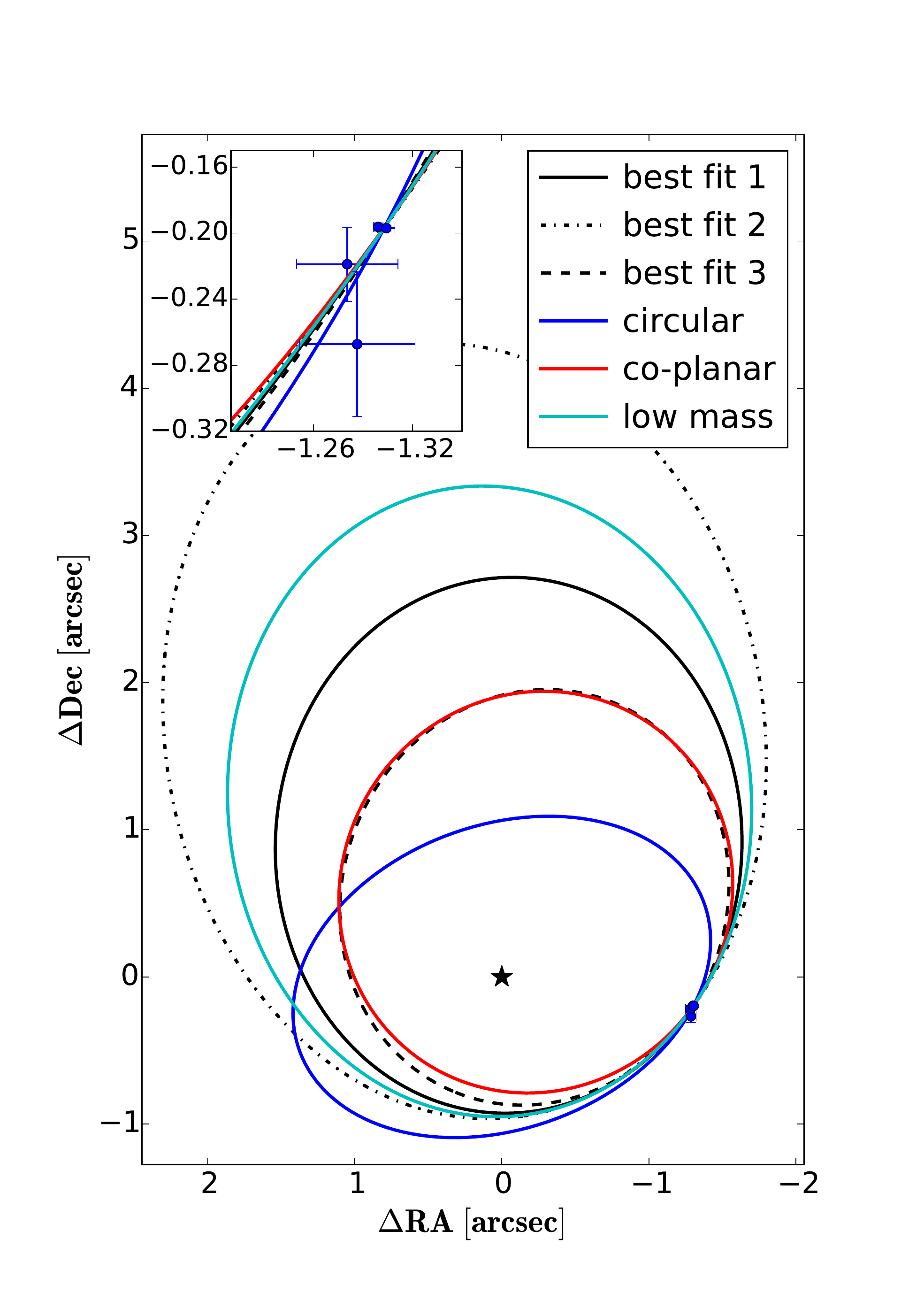}
\caption[]{Best fitting orbits to the current astrometry of the companion as recovered by our LSMC fit. The inset in the upper left is zoomed-in on the data points.}
\label{orbits}
\end{figure}

\emph{Photometry}\\
\\
In Fig.~\ref{sed} we show the measured SED of the companion, and compared it with theoretical models of low mass substellar objects calculated with the \texttt{Phoenix} (\citealt{2008ApJ...675L.105H}) atmosphere code using the AMES-Dusty models (\citealt{2000ApJ...542..464C}, \citealt{2001ApJ...556..357A}) for an age of 2\,Myr as input.
We tentatively explored a mass range between 2\,M$_\mathrm{Jup}$ and 20\,M$_\mathrm{Jup}$. The closest fit is achieved with a 5\,M$_\mathrm{Jup}$ planet corresponding to an effective temperature of 1700\,K.
However, it is clearly visible that even this best fit does not properly explain the measured photometry of the companion. 
While the J-H color may be explained by such an object (taking some red-ward shift due to extinction by circum-planetary material into account), the model clearly over-predicts the flux in K and L-band by an order of magnitude.
A significantly lower mass object of 2\,M$_\mathrm{Jup}$ corresponding to an effective temperature of 1100\,K could explain the K-band photometry, but still significantly over-predicts the L-band flux and is not compatible with the flux in the shorter wavelengths.
Generally there is no model that can explain all photometric data points and upper limits. 
This is a strong indication that we are looking at an object that is either significantly more complex than a "naked" planetary photosphere, or that the object (or the primary) is for some reason strongly variable.
Variability is indeed possible since all observation epochs have been taken months and sometimes decades apart.\\
One explanation for the peculiar shape of the SED could be a companion with a small (unresolved) surrounding disk. 
There are in fact two comparably faint objects in Chamaeleon known around which a circumplanetary disk is expected. One of them is the wide direct imaging companion to the young T Tau star CT\,Cha (\citealt{2008A&A...491..311S}).
CT\,Cha\,b shows Pa\,$\beta$ emission in the J-band (\citealt{2008A&A...491..311S}), as well as strong H\,$\alpha$ emission in the R-band (\citealt{2015ApJ...801....4W}), both strong indicators for ongoing accretion of material on the companion.
The companion mass is estimated to be 9-35\,M$_\mathrm{Jup}$ with a temperature range between 2500\,K and 2700\,K (\citealt{2008A&A...491..311S}, \citealt{2014A&A...562A.127B}, \citealt{2015ApJ...801....4W}).
We show the near infrared spectrum of CT\,Cha\,b along with optical photometry in Fig.~\ref{sed-small}. We overplot the photometry of the companion to CS\,Cha for comparison.
While R, I and H-band photometry are comparable in both objects, the J and K-band fluxes of CT\,Cha\,b are significantly larger than for the CS\,Cha companion. 
A second comparison object is the recently discovered free floating planetary mass object Cha\,J11110675-7636030 (\citealt{2017AJ....154...46E}), for which we also show available photometry in Fig.~\ref{sed-small}.
Assuming an age range of 1-3\,Myr and using a variety of planet evolutionary models, \cite{2017AJ....154...46E} find a mass range of 3-6\,M$_\mathrm{Jup}$ for this object. They note that the mid-IR photometry suggests the existence of excess emission best explained by circum-planetary material.
The object shows J-K colors similar to the CS\,Cha companion, and is also consistent with the L-band non-detection. The H-band photometry of both objects, on the other hand, differs significantly.
For both comparison objects CT\,Cha\,b and Cha\,J11110675-7636030 we have no information on the geometry of the surrounding circum-planetary material. In particular we do not know the inclination of these inferred disks.
It is possible that the companion around CS\,Cha is indeed more massive than both these objects, but is strongly extincted by a very inclined circum-companion disk. 
This scenario is indeed also supported by the high degree of linear polarization that we find for the companion. We thus explore several models with circum-companion material in section~\ref{model-section}.\\
\\
\emph{Astrometry}\\
\\
Since we have an observational baseline of $\sim$19\,yr, we attempted to fit the orbital motion of the companion around the primary stars. 
For this purpose we used the Least-Squares Monte-Carlo (LSMC) approach as described in \cite{2013MNRAS.434..671G}. 
We generated 10$^7$ random orbit solutions from uniform priors and then used these as starting points for a least squares minimization with the Levenberg-Marquardt algorithm.
In contrast to \cite{2013MNRAS.434..671G}, we did not assume a system mass but left it as a free parameter.
To limit the large parameter space we constrained the semi-major axis to values between 0.5\,arcsec and 3.0\,arcsec. 
This seems justified given the current position of the companion at $\sim$1.3\,arcsec and the fact that we see no significant change in separation between astrometric epochs.
In addition, we limited the total system mass to values between 0.9\,$M_\odot$ and 2.0\,$M_\odot$. 
The lower end of this mass interval is determined by the lower limit of the combined mass of the central binary star, i.e. in this case the companion mass would be small compared to the primary mass in the planet or brown dwarf regime.
The upper end is given by twice the upper limit of the central binary mass, i.e. in this case the companion would have roughly one solar mass.
We do not expect the companion to be more massive than the primary stars, since the resolved circumbinary disk would otherwise likely be truncated to an even smaller outer radius.\\
In Fig.~\ref{companion:lsmc} we show the resulting semi-major axis, inclination and mass versus eccentricity distributions of the 1\% best fitting orbits.
Since the uncertainties of the NACO and HST epochs are large compared to the SPHERE measurements, the fits are strongly dominated by the latter.\\
We find that the current astrometric epochs do not allow for constraint of the mass of the companion, since we find valid orbital solutions for the full range of input masses.
However, we can make a few observations about the system architecture. If the companion is indeed a Jovian planet or brown dwarf, then we can conclude that it must be on an eccentric orbit with the lower limit of the eccentricity between 0.2 and 0.26 depending on the central stars' masses.
In fact the total system mass should be above 1.4\,$M_\odot$ to allow for circular orbits. In this case the companion would be a low mass star with a mass between 0.4\,$M_\odot$  and 0.5\,$M_\odot$.
Independent of the mass, we find an upper limit for the eccentricity of 0.8. This upper limit is, however, introduced by our artificial cut off of the semi-major axis at 3\,arcsec. If we allow for larger semi-major axes, then we find even more eccentric orbits.
This correlation between semi-major axis and eccentricity is indeed common for orbits which are not well covered with observations (e.g. \citealt{2014MNRAS.444.2280G}).
Overall we find a peak of the eccentricity at $\sim$0.6. The vast majority of these eccentric orbits exhibits a face-on inclination. \\
It is interesting to investigate if co-planar orbits with the resolved circumbinary disk are possible since this could give an indication of the formation history.
We find that such co-planar orbits indeed exist. However, regardless of the total system mass there are no circular (e\,=\,0) co-planar orbits recovered.
Overall the distribution of the total mass and eccentricity closely match the non-coplanar case.\\
In Fig.~\ref{orbits} we show the three best fitting orbit solutions that were recovered by our LSMC fit as well as the best fitting solutions for a circular, co-planar and low mass (companion mass below 0.03\,M$_\odot$) orbit. The respective orbital elements are given in table~\ref{tab: orbit-elements}. The best fitting orbits are not co-planar and exhibit eccentricities between 0.41 and 0.63.
Since most of these orbits are seen face-on there would be a significant misalignment between the inclination of the resolved circumbinary disk and the orbital plane, as well as a misalignment with a putative highly inclined circum-companion disk.
Such spin-orbit and spin-spin misalignments in multiple systems are indeed predicted by hydrodynamic simulations of stellar formation in clusters (see e.g. \citealt{2012MNRAS.419.3115B, 2018MNRAS.475.5618B}) and were more recently observed in multiple systems with ALMA (see the case of IRAS\,43, \citealt{2016ApJ...830L..16B}). 
The total system masses for the best fitting orbits lie between 1.28\,$M_\odot$ and 1.84\,$M_\odot$, which puts the companion in the low stellar mass regime. However, we stress that lower (e.g. planetary) masses for the companion can not be ruled out with the existing astrometry. 
One example for an orbital solution that fits the astrometry and requires only a companion mass below 0.03\,M$_\odot$ is shown in Fig.~\ref{orbits}.
In general these best fitting orbits may still change significantly, with the availability of new high precision astrometric epochs in the future. Thus while the recovered distributions of orbital elements are meaningful,
we caution to over interpret these specific orbit solutions.\\
\\

\begin{table}
 \centering
 %\begin{minipage}{200mm}
  \caption{Orbit elements of the three best-fitting orbits shown in Fig.~\ref{orbits}, as well as the best fitting circular (c.), co-planar (c.-p.) and low mass (l.m.) orbits.}
  \begin{tabular}{@{}l@{\hspace{2pt}}cccccc@{}}
  \hline   
	\hline
 	 			& 1			&  2 			& 3 		& c.		& 	c.-p.		& l.m.	\\
 	\hline
	a\,["] 		& 1.82			& 2.65			& 1.43		& 1.46		& 1.48			& 2.14	\\
	m\,[M$_\odot$]	& 1.66			& 1.28			& 1.84		& 1.96		& 1.96			& 1.02	\\
	e			& 0.49			& 0.64			& 0.41		& 0		& 0.45			& 0.56						\\
	P\,[yr]			& 4048.4		& 8076.2		& 2667.0	& 2678.6	& 2731.6		& 6579.6						\\	
	i\,[deg]		& 0			& 0			& 0		& 45.2		& 21.6			& 0						\\
	$\Omega$\,[deg]		& 200.0			& 185.4			& 190.0		& 110.1		& 75.3			& 184.0						\\
	$\omega$\,[deg]		& 337.0			& 3.3			& 327.7		& 275.7		& 83.3			& 0						\\
	T$_0$\,[yr]		& 1651.9		& 1668.1		& 1602.1	& 3011.5	& 4363.7		& 1593.6						\\
 \hline\end{tabular}

\label{tab: orbit-elements}
\end{table}

\subsection{Detection of a circum-companion disk?}
\label{model-section}

To test whether the peculiar measurements of the companion can be explained by a substellar object surrounded by a disk, we aimed to model its photometric (SED) and polarimetric (degree of polarization) properties using the radiative transfer code RADMC3D. In all our models, we consider astronomical silicates (\citealt{2001ApJ...548..296W}), and use a single dust size for simplicity. 
We again consider the AMES-DUSTY atmosphere models as input for the central object.
We run 3 different families of model. 

\begin{figure*}
\centering
\includegraphics[width=0.98\textwidth]{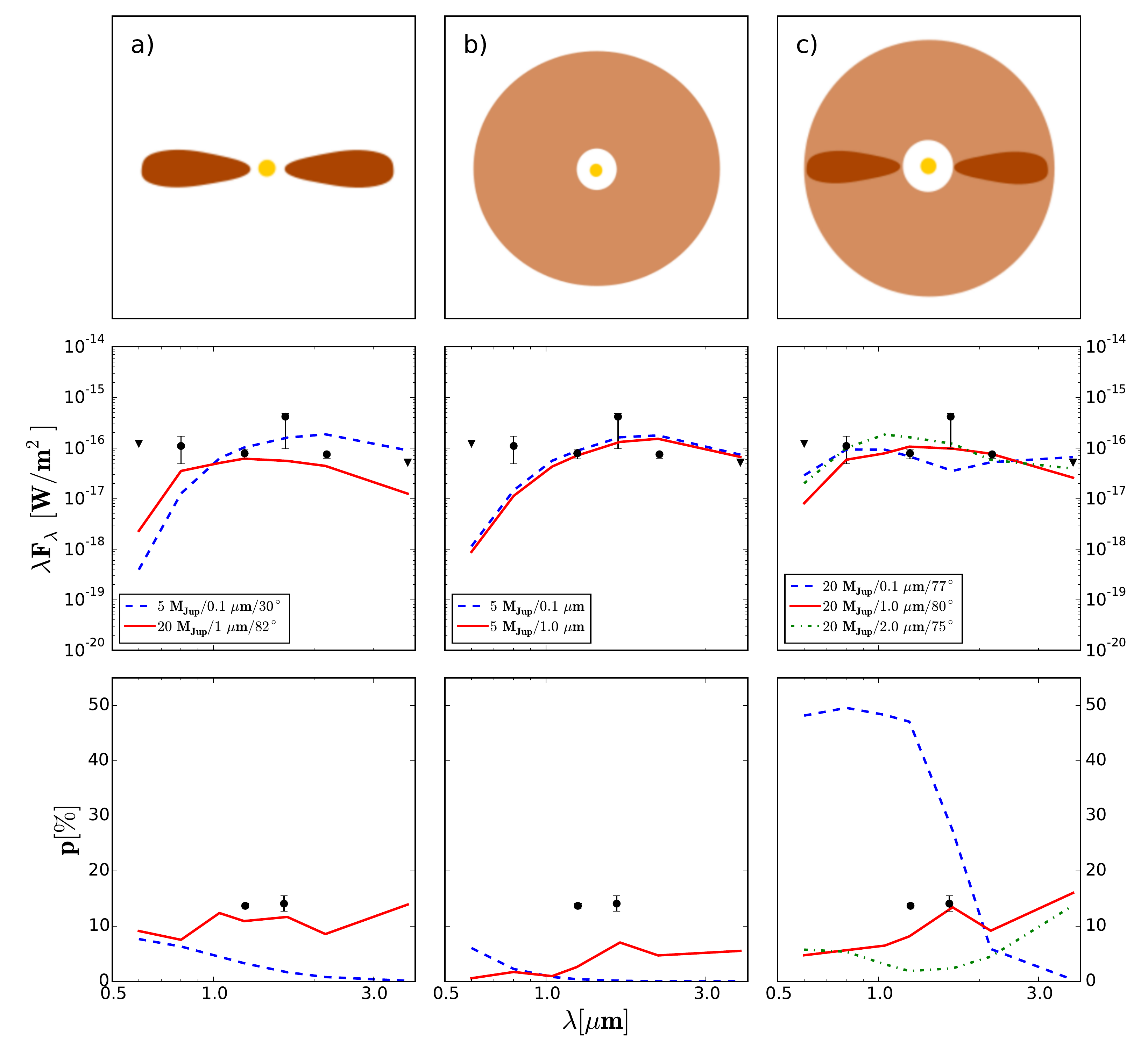}
\caption[]{\textit{1st row:} Sketches of the model families described in section \ref{model-section}. We show in all cases a cross-section. \textit{2nd row:} Photometry of the companion along with the model photometry for different model parameters. The legend gives information about the assumed companion mass, the size of the considered dust grains, as well as the circumplanetary disk inclination (in models a) and c)). 
\textit{3rd row:} Same as the second row, but for degree of linear polarization. Colors and line styles represent the same models as in the previous row.}
%\end{sidecaption}
\label{models}
\end{figure*}

(a) Our first model includes a disk around a substellar companion. The circumplanetary disk extends up to 2 au. This maximum outer radius was inferred from the fact that we do not resolve the companion. Its density is described as:
\begin{equation}
\rho_{\rm{disk}} (r,z) = \frac{\Sigma(r)}{\sqrt{2\pi}H_{\rm{p}}(r)} \exp\left(\frac{-z^{2}}{2H_{\rm{p}}(r)^{2}}\right)
\end{equation}
where $\Sigma(r)$ is the surface density and $H_{\rm{p}}(r)$ is the pressure scale height. Both quantities are described as power laws with radius, with exponents $\zeta$ (flaring) and $p$. 
The model parameters are given in Table~\ref{tab:modelparam}.
The model has a complex parameter space, which we explored qualitatively by varying the mass and thus the luminosity of the central object, the grain size, the disk mass and inclination.
To produce a significant level of polarization (above 5\%), the disk must be strongly inclined (seen close to edge-on), in turn extinguishing the thermal emission from the companion and the innermost disk radii. 
To match the SED at high inclinations we thus must increase the mass of the central object. An increase in disk mass has an effect similar to an increase of the inclination on the SED of the central object.
Finally the grain size can be varied to modulate the polarization efficiency. We considered grain sizes of 0.5\,$\mu m$, 1\,$\mu m$ and 2\,$\mu m$.
We show a sketch of the model, along with the best fitting results for a 5\,M$_\mathrm{Jup}$ and a 20\,M$_\mathrm{Jup}$ companion in the left column of Fig.~\ref{models}.
For the lower mass we require a low disk inclination in order to get enough flux of the companion in the near-infrared. However, this model under-predicts the I-band flux and over-predicts the L-band flux, in addition to not revealing significant polarization. 
For the higher mass we find a much better fit. We can increase the disk inclination to much higher values, which match the J and H-band polarization well.
Furthermore, the resulting SED is a close fit to all photometric measurements of the companion, excluding the H-band.
We have also investigated whether we can derive an upper mass limit for the companion by placing a 72\,M$_\mathrm{Jup}$ companion in the center of the disk. 
We found that even for high inclinations, such a model severely over-predicts the K and L-band flux.\\
Note that in this model, we do not consider the binary as a source of irradiation. 
The only way that the companion would scatter a significant amount of light from the central binary, at that distance, would be if it stands outside of the plane of the circumbinary disk. Otherwise the light from the central binary would be blocked by the disk.  
We therefore tested the same model, but with an additional irradiation source (the central binary) at 214\,au, and after placing the companion and disk outside of the plane of the circumbinary disk. 
We find that the scattered light signal from the central binary alone is between 2-3 orders of magnitude fainter than our measurements (see Fig.~\ref{app: binary_irradiation} for comparison). 
It thus only has a marginal influence on our modeling results and was ignored for simplicity.

(b) We then changed our models to test a different geometry, and considered an envelope of dust grains surrounding the companion, as this should enhance the amount of scattered light. The density structure is given by: 
\begin{equation}
\rho_{\rm{env}} (r,z) = \rho_{0}\left( \frac{\sqrt{r^{2} + z^{2}}}{  \sqrt{R_{\rm{out}}^{2} + z_{\rm{out}}^{2}}} \right) ^{q}
\end{equation}
where  $\rho_{0}$ is the density of the envelope at its outer radius. We chose the mass of the envelope so that its optical depth would be the same as in the disk model.
Since in this model we do not have a disk to modulate the flux of the companion, we only tested models for a 5\,M$_\mathrm{Jup}$ central object, which provided the closest match to the companion SED.
The results are shown in the middle column of Fig.~\ref{models}.
We find a similar match to the SED as in the disk model for a low mass companion, but our models underestimate the degree of polarization with values at most in the order of a few \% ($\sim$7\% for micron-size dust grains). 

(c) Our final model is a combination of the previous two. 
We consider a companion surrounded by a disk plus an additional envelope. 
For this model, we consider a 20\,M$_\mathrm{Jup}$ companion since it provided the best fit for the disk-only model.
The density at each (r,z) is taken as the maximum between $\rho_{\rm{disk}}$ and $\rho_{\rm{env}}$. 
Note that the mass of the envelope is negligible compared to that of the disk ($\sim$0.4\%). 
This configuration allows us to obtain a large degree of polarization, by increasing the amount of scattered light with the envelope, while reducing the total intensity from the central object with an inclined disk. 
In the right column of Fig.~\ref{models}, we show the results for three models with different grain sizes. Note that we also varied the inclination of the disk in order to obtain a good fit of the data.
Although none of our models fit both the photometry and the level of polarization perfectly, we find that a disk composed of 1\,$\mu$m sized dust grains and a high inclination of 80$^\circ$ is consistent with the observed photometry.
This model still under predicts the polarization in the J-band by a factor of $\sim$1.7. Our model using smaller grains on the other hand will fit the HST and SPHERE J-band photometry slightly better, while it misses the SPHERE H-band and NACO L-band measurements.
Smaller grains also lead to a dramatic over prediction of the degree of polarization in the near infrared. Larger grains than 1\,$\mu$m do not significantly contribute to the degree of polarization in the J and H-bands.
Given these results it is conceivable that a more complex grain size distribution (instead of a single grain size) including grain sizes between  0.1\,$\mu$m and 1\,$\mu$m may be able to reproduce the degree of polarization as well as the photometry.
However, we would like to point out that the parameter space is complex and degenerate between multiple parameters, such as companion mass, disk inclination and dust grain size.
Thus we do not claim that the disk plus dust envelope model with the given parameters is the only model that can reproduce our measurements. 
Additional measurements are needed before an attempt is made to constrain the nature of the companion to CS\,Cha further. 
An observation with SPHERE/ZIMPOL to detect the companion in optical polarized light could help to constrain the dust grain sizes as well as the presence of a dust envelope.
An ALMA observation on the other hand may constrain the mm-dust mass at the companion position and thus indirectly the mass of the companion itself.\\
From the angle of polarization we can deduce the geometry of such a system. The angle of polarization will mostly be determined by the region of the unresolved disk from which we receive the largest amount of polarized light.
In the disk only model, this is the earth-facing forward scattering side of the disk, and in the the disk+envelope model these are the "poles" of the circular envelope away from the disk.
In both cases we would thus expect that the angle of polarization is aligned or closely aligned with the position angle of the circum-companion disk. 
We note that this scenario would change in the presence of an outflow which dominates as the source of scattered light. In such a case we would expect the angle of polarization to be perpendicular to the disk plane (\citealt{1989AJ.....98.1368T}). However, we have not modelled such a scenario.
We have in general, not included this geometrical consideration in our models since the degree of polarization and the photometry are independent of the disk position angle.\\

\begin{table}[!h]
	\begin{center}
	\caption{Radiative transfer model parameters.}
	 \begin{tabular}{lccc}
	 	 \hline
	 	 \hline
		 Model 	 &  (a) Disk & (b) Envelope & (c) Disk  	\\
		 Parameters &  & & \& Envelope \\
	 	 \hline
 		M$_{\rm{comp}}$ [M$_{\rm{Jup}}$]& 5/20  & 5 & 20 \\
		T$_{\rm{eff}}$ [K] & 1580/2500 & 1580 & 2500 \\
		R$_{\rm{comp}}$ [R$_{\odot}$] & 0.17/0.25 & 0.17 & 0.25\\
		R$_{\rm{in}}$  [au] & 0.003 & 0.003 & 0.003 \\
		R$_{\rm{out}}$ [au] & 2 & 2 & 2 \\
		M$_{\rm{disk}}$ [M$_{\odot}$]& 1.9 10$^{-7}$  & - & 1.9 10$^{-7}$ \\
		M$_{\rm{env}}$ [M$_{\odot}$]& - & 8.5 10$^{-9}$ & 2.3  10$^{-10}$ \\
		$\rho_{0}$ [g/cm$^{3}$]& - & 1 10$^{-16}$  &  5 10$^{-17}$\\	
		H$_{\rm{p}}$(R$_{\rm{out}}$)/R$_{\rm{out}}$ & 0.18 & - & 0.18 \\
		$\zeta$ & 0.25 & - & 0.25 \\
		$p$  & -1 & - & -1 \\
		$q$  & - & -1 & -1 \\		
 		 \hline
 		 \end{tabular}
 	\label{tab:modelparam}
 	\end{center}
	%\tablecomments{}
	
\end{table}

\section{Summary and conclusions}

We observed the CS\,Cha system for the first time in high resolution polarimetry with SPHERE/IRDIS in J and H-band.
We resolved a circumbinary disk with an outer extent of 55.6\,au in scattered light. The disk cavity predicted by previous studies was not detected due to the limited inner working angle of the coronagraph.
The upper limit for the radius of the disk cavity is 15.3\,au, consistent with previous models by \cite{2016MNRAS.458.1029R} using unresolved Herschel data.
We find that the disk has an inclination of 24.2$^\circ \pm$3.1$^\circ$.\\
Outside of the disk at a projected separation of 214\,au we find a faint companion with an extreme degree of linear polarization. 
To our knowledge this is the first faint and likely very low mass companion to a nearby star that has been discovered in polarized light.
With HST and NACO archival data we show with high confidence that the companion is co-moving with the primary stars, and is thus bound to the system, placing it at the same distance and age as CS\,Cha.
The complex photometry of the companion could not be explained with current atmosphere models. If just J and H-band were considered, a 5\,M$_\mathrm{Jup}$ mass may be inferred.
However, this does not fit the photometry in other bands, in particular the non-detection in L-band.
Furthermore, a "naked" substellar companion is expected to have a low intrinsic polarization. 
\cite{2017arXiv170609427S} showed recently that the expected degree of linear polarization from such a companion due to rotational oblateness or patchy cloud covers should not exceed 3\,\% and is typically lower\footnote{See also the previous work by \cite{2001ApJ...561L.123S}, who find a similar range for the degree of linear polarization.}.
Thus we suggest that we are looking at a companion with a surrounding disk or dust-envelope.
We explored the wide parameter space for such a model with the radiative transfer code RADMC3D. 
We find that we can explain the companion SED and polarization reasonably well, with either a highly inclined disk around a 20\,M$_\mathrm{Jup}$ object or with a disk and additional dust envelope around an object of the same mass.
This puts the companion clearly in the substellar regime: either a very low mass brown dwarf or a high mass planet.\\
From our orbit fit to the available astrometry over a time base line of 19\,yr, we can conclude that the orbit of the companion is likely eccentric with a minimum eccentricity of 0.3.
This gives some indication of how the companion may have formed. For an in-situ formation, either by core-accretion or by gravitational collapse in the outer circumbinary disk, one would not expect an eccentric orbit.
Also the strong misalignment of the circumbinary and the circum-companion disk do not fit these scenarios. 
However, the eccentricity may be explained by dynamical interaction with the unresolved stellar binary. 
The two systems could be caught in Kozai-Lidov type resonances effectively exchanging relative inclination and eccentricity (see e.g. \citealt{2005ApJ...627.1001T}).
Another possibility for an eccentric orbit would be the formation at close separations in the circumbinary disk and a subsequent dynamical scattering event in which again the central binary may have played a role.
However, in such a scenario one would expect that the companion lost the surrounding disk and that some sign of perturbation would be visible in the circumbinary disk. Both do not seem to be the case.\\
While for typical planet formation scenarios the location and eccentricity of the orbit of the CS\,Cha companion are problematic, this is less so for a more star-like formation by collapse in the molecular cloud in which also the CS\,Cha binary formed.
In such a case the misaligned disks around the companion and the stellar sources would also not be problematic as many such examples are known, most prominently the HK\,Tau system (\citealt{1998ApJ...502L..65S}) with a similar configuration as the CS\,Cha system.\\
To better constrain the mass and properties of the companion and its surrounding disk, additional observational data is necessary, in particular, ALMA observations will allow to detect the amount of mm-sized dust around the companion, and likely, reveal its true nature.
Additional SPHERE/ZIMPOL observations that would help to determine the grain size distribution and also potentially if the disk or disk plus envelope scenario explains the system configuration best. \\
Only few other systems are known that harbor a sub-stellar companion with a disk around it, such as the FW\,Tau (\citealt{2014ApJ...781...20K, 2015ApJ...798L..23K}) system or the 1SWASP\,J140747.93-394542.6 system (\citealt{2012AJ....143...72M}, \citealt{2015MNRAS.446..411K}).
The former confirmed by ALMA observations, while the latter was detected in transit.
However, the CS\,Cha system is the only systems in which a circumplanetary disk is likely present as well as a resolved circumstellar disk. 
It is also to the best of our knowledge the first circumplanetary disk directly detected around a sub-stellar companion in polarized light, constraining its geometry.
Once the system is well understood it might be considered a benchmark system for planet and brown dwarf formation scenarios.

\begin{acknowledgements}
We thank an anonymous referee for significantly improving our original manuscript.
We acknowledge I. Pascucci, M. Min, M. Hogerhijde, C. Dominik and G. Muro-Arena for interesting discussions. 
MB acknowledges funding from ANR of France under contract number ANR-16-CE31-0013 (Planet Forming Disks).
AJ acknowledges the support by the DISCSIM project, grant agreement 341137 funded by the European Research Council under ERC-2013-ADG.
JO acknowledges support from the Universidad de Valpara\'iso and from ICM N\'ucleo Milenio de Formaci\'on Planetaria, NPF.
The research of FS leading to these results has received funding from the European Research Council under ERC Starting Grant agreement 678194 (FALCONER).
SPHERE is an instrument designed and built by a consortium consisting of IPAG (Grenoble, France), MPIA (Heidelberg, Germany), LAM (Marseille, France), LESIA (Paris, France), Laboratoire Lagrange (Nice, France), INAF - Osservatorio di Padova (Italy), Observatoire de Geneve (Switzerland), ETH Zurich (Switzerland), NOVA (Netherlands), ONERA (France) and ASTRON (Netherlands) in collaboration with ESO. SPHERE was funded by ESO, with additional contributions from CNRS (France), MPIA (Germany), INAF (Italy), FINES (Switzerland) and NOVA (Netherlands). SPHERE also received funding from the European Commission Sixth and Seventh Framework Programmes as part of the Optical Infrared Coordination Network for Astronomy (OPTICON) under grant number RII3-Ct-2004-001566 for FP6 (2004-2008), grant number 226604 for FP7 (2009-2012) and grant number 312430 for FP7 (2013-2016). 
This research has made use of the SIMBAD database as well as the VizieR catalogue access tool, operated at CDS, Strasbourg, France. This research has made use of NASA's Astrophysics Data System Bibliographic Services. 
Finally CG would like to thank Donna Keeley for language editing of the manuscript.
\end{acknowledgements}


\bibliographystyle{aa} % style aa.bst
\bibliography{myBib} % your references Yourfile.bib

%\Online

\begin{appendix} %First online appendix

\section{Stellar PSF in the SPHERE polarimetric images}

\begin{figure}[!h]
\centering
\includegraphics[width=0.95\columnwidth]{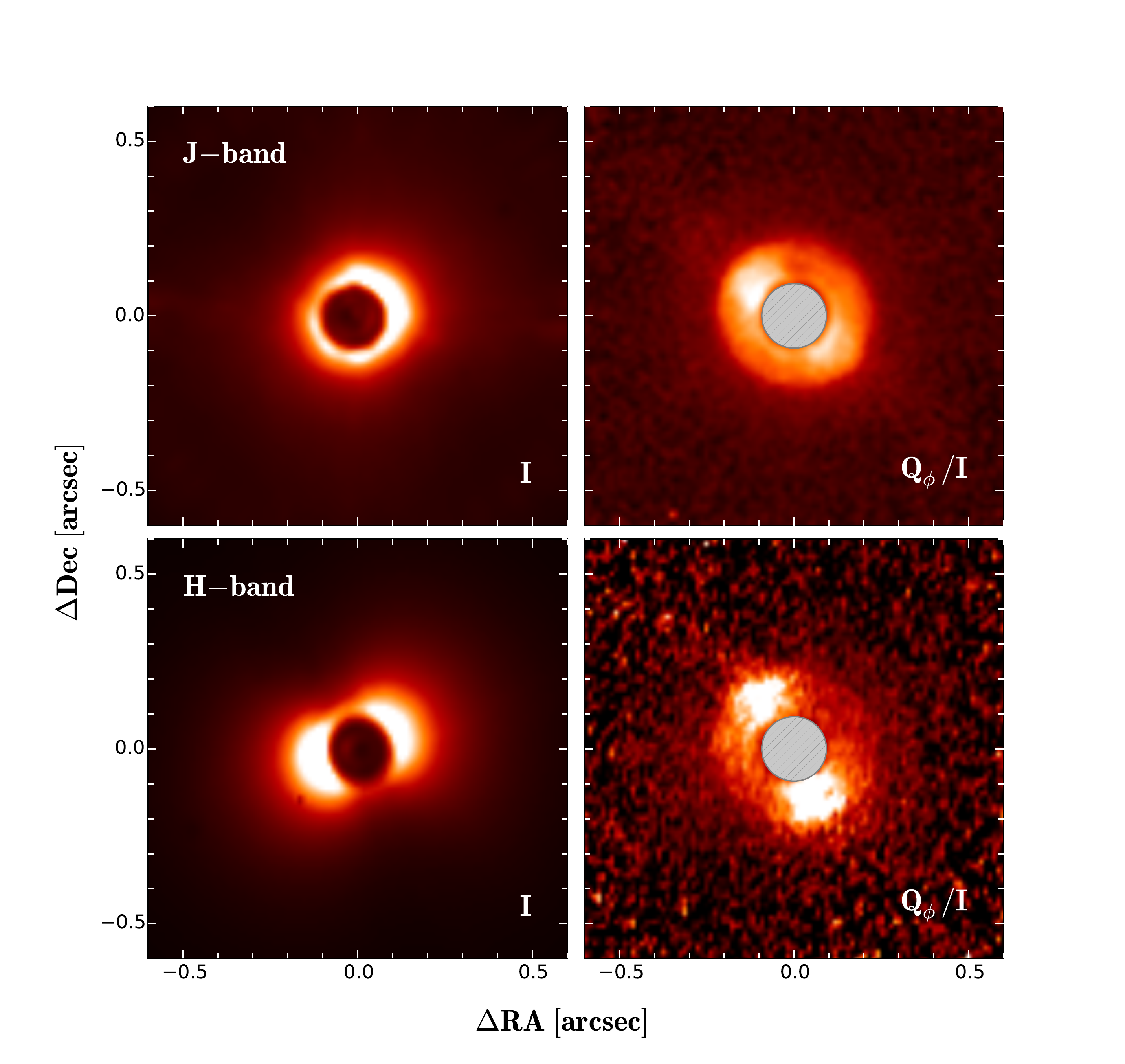}
\caption{\textit{Left Column:} Total intensity images derived from our polarimetric observations with SPHERE in J and H-band. The circumstellar disk is not visible in these images since flux is completely dominated by the bright stellar PSF.
The visible "cavity" in the image center is caused by the SPHERE/IRDIS coronagraph. The stellar PSF is strongly elongated in the South-East to North-West direction in the H-band observations. Additionally the star is not well centered behind the coronagraph, as is visible in the illumination pattern behind the (slightly transparent) coronagraph.
\textit{Right Column:} Q$_\phi$ images from Figure~\ref{disk-main} divided by the total intensity images in the left column. Note that this is not equivalent with polarization degree, since the intensity images are dominated by the stellar PSF rather than disk signal.
The visible bright lobes are aligned with the semi-minor axis of the stellar PSF, i.e. they are caused by a drop off in stellar flux and not by the polarized phase function of the disk scattered light.
It is visible that the assymmetric scattered light signal in the H-band is less apparent in this image, indicating that it originates from the asymmetric flux in the stellar PSF which was not well centered behind the coronagraph, rather than from an actual astrophysical asymmetry in scattered light from the disk.}
\label{app: sphere_stellar}
\end{figure}

\section{Color comparison of the companion with YSOs in Cha}

\begin{figure}[!h]
\centering
\includegraphics[width=0.95\columnwidth]{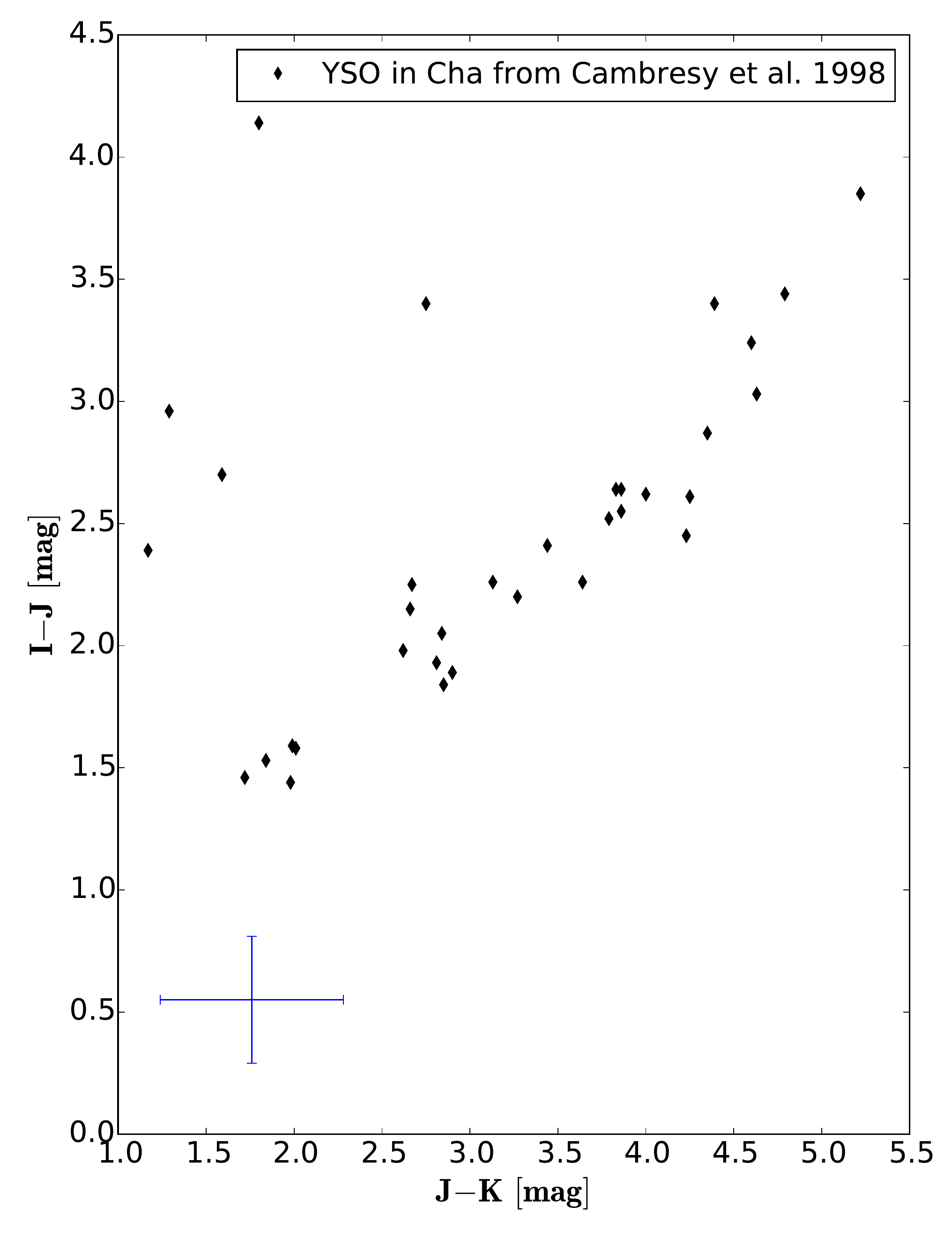}
\caption{Color-color plot of YSOs in Cha from \cite{1998A&A...338..977C}. For comparison we added the companion to CS\,Cha in the plot as blue cross (marking the uncertainties of the photometry).
The companions colors are much bluer in I-J than expected for a YSO. In J-K they are still very blue compared to most objects. 
Overall we can conclude that the companion does not match YSO colors well and is in all likelihood no embedded background YSO.}
\label{app: cambresy}
\end{figure}

\section{Model of a circum-companion disk irradiated by the central binary star}

\begin{figure}[!h]
\centering
\includegraphics[width=0.95\columnwidth]{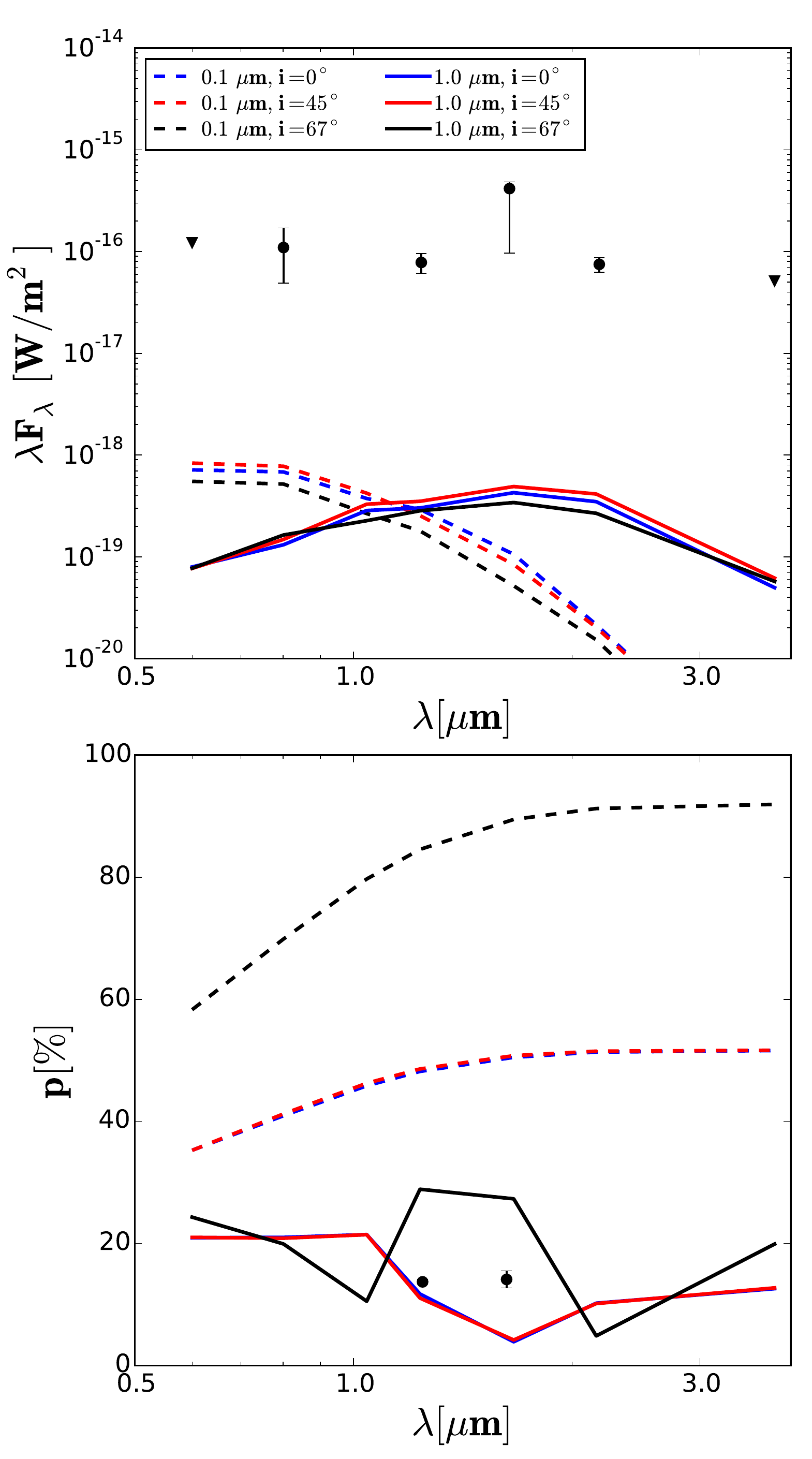}
\caption{Photometry and degree of linear polarization of the companion assuming the central object is of low enough mass that its thermal emission is negligible.
This very low mass object is still surrounded by a circum-companion disk.
Thus the light received from the companion is entirely scattered light from the primary stars. 
We used different grain sizes and inclinations of the circum-companion disk, indicated by line style and color. 
We find that, while we can explain the degree of linear polarization with such a model, the received flux is several order of magnitude below our measurements.}
\label{app: binary_irradiation}
\end{figure}

\end{appendix}

\end{document}